\documentclass{aa}

\usepackage{graphicx}
\usepackage{txfonts}
\usepackage{gensymb}
\usepackage[hidelinks]{hyperref}

\newcommand{\doublehline}{\hline\hline\\[-1.5ex]}

\begin{document}

   \title{Magnetic properties of the Abell 3391-3395 system revealed using wide-field MeerKAT polarimetry}
\titlerunning{Magnetic properties of the Abell 3391-3395 system revealed using wide-field MeerKAT polarimetry}

   \subtitle{}

   \author{V. Gustafsson
          \inst{1}
          \and
          M. Br\"uggen \inst{1}
          \and
          C. Tasse \inst{2,3}
          \and
          S. P. O'Sullivan \inst{4}
          }

   \institute{$^1$ Hamburger Sternwarte, University of Hamburg, Gojenbergsweg 112, 21029 Hamburg, Germany\\
   $^2$ GEPI, Observatoire de Paris, Université PSL, CNRS, 5 Place Jules Janssen, 92190 Meudon, France\\
   $^3$ Department of Physics \& Electronics, Rhodes University, PO Box 94, Grahamstown 6140, South Africa\\
   $^4$ Departamento de Física de la Tierra y Astrofísica \& IPARCOS-UCM, Universidad Complutense de Madrid, 28040 Madrid, Spain
   }

   \date{Received 7 April 2026 / Accepted 25 July 2026}

\abstract
{Magnetic fields in cluster outskirts and the intercluster medium remain poorly constrained because diffuse synchrotron emission is difficult to detect owing to its low surface brightness. Faraday rotation measures (RMs) of polarized background sources can be used to probe large-scale foreground structures. The nearby interacting Abell 3391-3395 system hosts a well-established X-ray bridge, making it a prime target to investigate magnetization in the intercluster environment.}
{We aim to characterize the magnetized environment of Abell 3391/95 and its surroundings by constructing a dense RM grid from wide-field polarimetry.}
{We observed Abell 3391/95 with MeerKAT in full polarization using a three-pointing mosaic. The data were calibrated with direction-independent and direction-dependent techniques and imaged using visibility plane mosaicing to obtain sensitive coverage across the field. Using Faraday synthesis, we formed Faraday cubes and measured RMs for polarized background sources. We defined on- and off-target regions using contours from a wavelet-filtered eROSITA image.}
{We identified a total of 434 polarized sources, with a polarized source density ranging from about 30 sources per square degree in the outer regions to about 110 sources per square degree in the central part of the field, and a field-averaged density of 73 sources per square degree. The clusters show a statistically significant enhancement of RM scatter relative to the off-target region. In contrast, the bridge between the clusters shows a comparatively low RM scatter, while an RM structure-function analysis on matched angular scales yields a tentative indication of larger RM differences in the bridge than off-target. Combined with the low per-source depolarization, this points to a magnetic field in the bridge that is relatively ordered on scales on the order of 10 kpc, but less ordered on larger scales. The non-detection of diffuse synchrotron emission in the bridge yields much improved upper limits on the emissivity.}
{}

\keywords{}

    \maketitle

\section{Introduction}
\label{sec:introduction}

Magnetic fields are thought to pervade the cosmic web, influencing the dynamics and long-term evolution of baryons over a wide range of physical scales \citep[e.g.,][]{donnert_2018}. Within galaxy clusters, observations routinely reveal microgauss-level fields permeating the intracluster medium (ICM), detected through diffuse synchrotron radiation and Faraday rotation of embedded and background radio sources \citep[e.g.,][]{govoni_2004,stuardi_2021,loi_2025}. These fields are expected to affect key physical processes in the ICM, including the regulation of anisotropic thermal conduction, the generation and decay of turbulence, magnetic draping at cold fronts, and the acceleration and transport of cosmic rays \citep[e.g.,][]{zuhone_2013,dursi_2008,brunetti_2011}.

In the ICM, the amplification of magnetic fields from weak seeds is commonly attributed to a combination of adiabatic compression and small-scale dynamo action driven by mergers and gas motions \citep[e.g.,][]{vazza_2018,brunetti_2014}. While these mechanisms can account for the microgauss field strengths inferred in many systems, the behavior and organization of magnetic fields in lower-density environments such as cluster outskirts and the intercluster medium remain considerably less well constrained, both observationally and theoretically \citep[e.g.,][]{locatelli_2021,carretti_2025}. Establishing the magnetization in these regimes is critical because it encodes information about seeding channels, subsequent amplification, and the coupling between magnetic fields, turbulence, and the thermal plasma during structure formation.

Outside massive cluster interiors, cosmological simulations predict that a large fraction of baryons reside in the warm-hot intergalactic medium (WHIM) within filaments that connect groups and clusters of galaxies \citep[e.g.,][]{cen_1999,shull_2012,tuominen_2021}. Compared to cluster cores, these regions have much lower electron densities and are expected to host substantially weaker magnetic fields, potentially reflecting the seeding and amplification history of the cosmic web \citep[e.g.,][]{locatelli_2021,carretti_2025}. Direct observational constraints in this regime are scarce. Deep radio observations have revealed shocks and radio relics tracing energetic processes at the peripheries of clusters \citep[e.g.,][]{vanweeren_2019,degasperin_2022}, but the detection of diffuse synchrotron emission associated with filaments remains rare on an object-by-object basis. Statistical approaches, including stacking analyses and searches for systematic excesses, provide growing evidence for magnetized plasma on megaparsec scales \citep[e.g.,][]{vernstrom_2021,vernstrom_2023}, yet robust measurements for individual filaments are still limited.

Faraday rotation offers a complementary way to probe magnetization in low-density environments. It rotates the position angle of a linearly polarized signal by an amount $\Delta \chi$ proportional to the wavelength squared $\lambda^2$:

\begin{equation}
\label{eq:rotation}
    \Delta \chi = \phi \lambda^2,
\end{equation}

where $\phi$ is the Faraday depth, defined as

\begin{equation}
\label{eq:Faraday_depth}
    \phi(r) = \big(0.81\;\text{rad m$^{-2}$}\big)\int_0^r \frac{\text{d}r'}{\text{pc}}\frac{n_e(r')}{\text{cm}^{-3}}\frac{B_r(r')}{\mu \text{G}},
\end{equation}

where $n_e$ is the number density of thermal electrons, and $B_r$ is the line-of-sight component of the magnetic field. Modern broadband techniques such as rotation measure (RM) synthesis enable the sensitive recovery of Faraday components even in complex fields \citep{brentjens_2005}, by decomposing the polarized broadband signal into contributions at different Faraday depths. 

Dense RM grids of extragalactic background sources provide a statistical means to map magnetized plasma in the foreground, even when diffuse synchrotron emission is too faint to detect directly \citep[e.g.,][]{gaensler_2025,anderson_2024}. Early results from wide-area surveys demonstrate the potential of this approach. The Polarisation Sky Survey of the Universe's Magnetism (POSSUM) pilot survey has delivered RM source densities on the order of $30$ to $40~\mathrm{deg}^{-2}$ \citep{vanderwoude_2024}, and recent work has used these dense grids to identify enhanced RM scatter and coherent RM trends associated with supercluster-scale environments \citep[e.g.,][]{lopez_2025}. At lower frequencies, the Low-Frequency Array (LOFAR) Two-metre Sky Survey Data Release 2 (LoTSS DR2) RM grid provides a complementary catalog of Faraday depths over large areas of sky \citep{osullivan_2023}, as well as allowing analyses of LOFAR-based RM samples to place new constraints on magnetized filaments and on magnetogenesis scenarios \citep[e.g.,][]{carretti_2025}.

The Abell 3391 (A3391) and Abell 3395 (A3395) (collectively Abell 3391/95) system at a redshift of $z \approx 0.053$ provides a particularly promising environment in which RM grid techniques can be applied. The system consists of two interacting galaxy clusters connected by a bridge of warm gas that has been studied in X-rays for more than two decades \citep{tittley_2001}. Observations with eROSITA, together with earlier X-ray measurements, reveal a bridge extending several megaparsecs between the clusters and an additional structure suggestive of ongoing accretion from the surrounding large-scale environment \citep{reiprich_2021,veronica_2022,alvarez_2018,sugawara_2017}. Thermal Sunyaev-Zel'dovich measurements further support the presence of diffuse plasma connecting the pair, consistent with a premerger configuration in which gas is being assembled in the intercluster region \citep{reiprich_2021}. Despite this rich structure, radio observations of A3391/95 remain limited. A wide-field continuum study with the Australian Square Kilometre Array Pathfinder (ASKAP) reported no convincing evidence for diffuse synchrotron emission associated with the intercluster bridge, reaching surface-brightness sensitivities of approximately $25$ to $35~\mu\mathrm{Jy,beam^{-1}}$ at an angular resolution of about $10\arcsec$ \citep{bruggen_2021}. 

\begin{figure*}
    \centering
    \includegraphics[width=\textwidth]{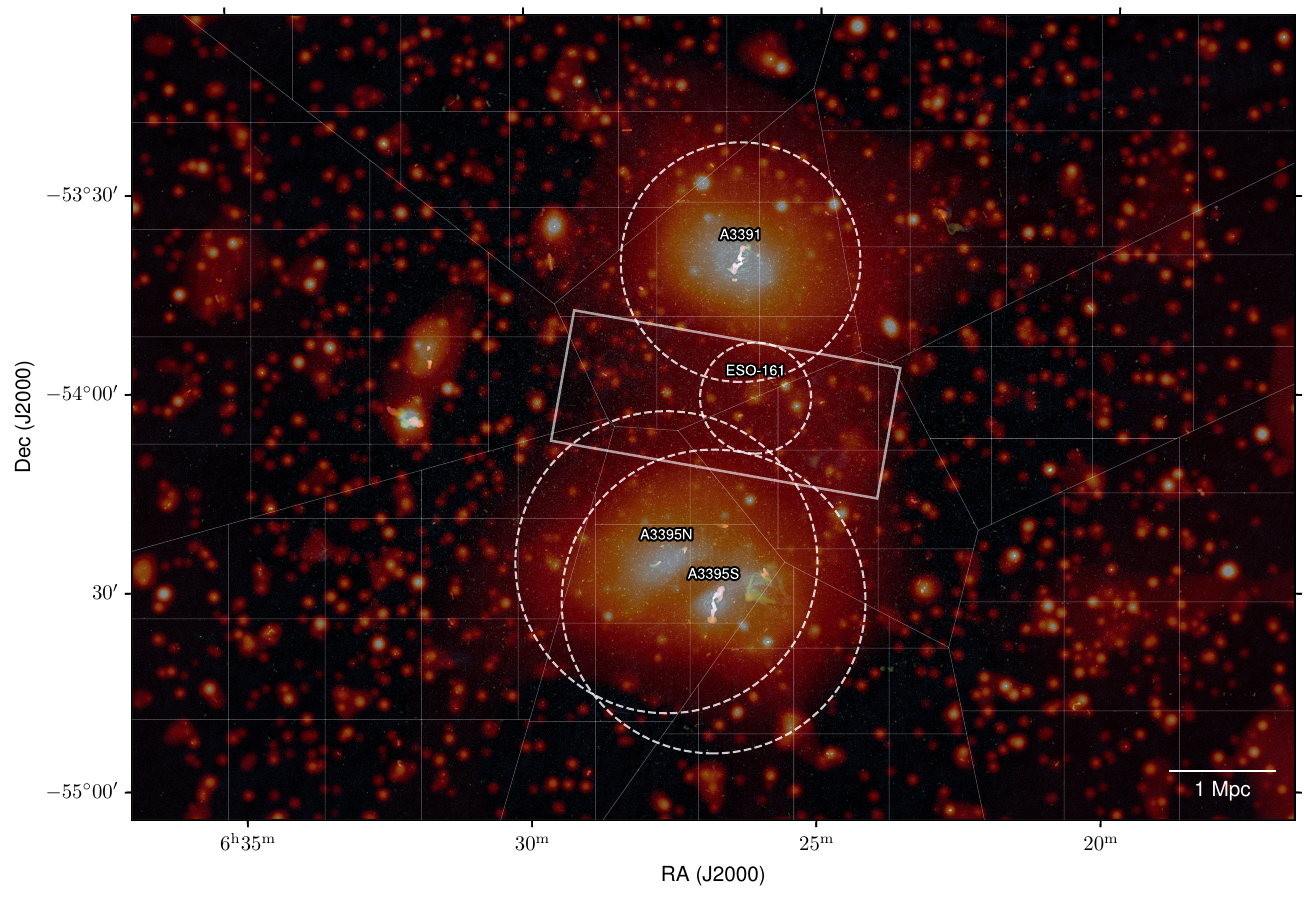}
    \caption{Stokes~$I$ radio mosaic overlaid on the eROSITA wavelet-filtered X-ray image. Compact sources correspond to radio continuum emission and are shown with a cubehelix color scale, whereas the extended diffuse background traces the X-ray surface brightness and is shown in red-to-white tones. White dashed circles represent the $r_{500}$ radii of the galaxy clusters A3391, A3395S, and A3395N, and the galaxy group ESO-161. The white solid rectangle defines the region used for the bridge analysis. The faint white lines show the faceting and tessellation used in direction-dependent imaging and calibration. The scale bar in the bottom right shows the physical scale at the mean cluster redshift of $z=0.053$, assuming a flat $\Lambda$CDM cosmology with $H_0 = 70~{\rm km~s^{-1}~Mpc^{-1}}$ and $\Omega_{\rm m}=0.3$. At this redshift, 1 Mpc corresponds to 16.1 arcmin.}
    \label{fig:x_ray}
\end{figure*}

MeerKAT can offer a deeper view of this system. Its large collecting area and compact core provide excellent surface-brightness sensitivity to emission on arcminute scales, making it well suited to detect faint extended synchrotron structures \citep{jonas_2016}. Moreover, MeerKAT's full polarization capability enables the construction of a dense RM grid across a wide field, allowing magnetized plasma to be traced statistically even when diffuse emission remains undetectable. In this work, we observe the A3391/95 system with MeerKAT in full polarization and apply an advanced reduction strategy, including direction-dependent calibration, visibility plane mosaicing, and wide-field polarimetric imaging, to reach substantially improved sensitivity. This allows us to investigate both total-intensity and Faraday-rotation signatures across the full system, from cluster interiors and outskirts to the intercluster environment.\\

This paper is organized as follows. In Sec.~\ref{sec:observation} we present the X-ray and radio observations used in this work. We also describe the MeerKAT observing setup and outline the calibration strategy, from cross-calibration through self-calibration to direction-dependent solutions, including our use of visibility plane mosaicing for wide-field imaging. In Sec.~\ref{sec:final_imaging} we present the final total-intensity products and the polarimetric imaging, culminating in the construction of a linearly polarized Faraday depth cube. In Sec.~\ref{sec:rm_grid} we describe the polarized-source cataloging used to construct the RM grid, apply Galactic-foreground corrections, and present the resulting RM statistics across the A3391/95 system and its intercluster environment. In Sec.~\ref{sec:discussion} we briefly discuss our results, before we conclude in Sec.~\ref{sec:conclusions}.

\section{Observations and data reduction}
\label{sec:observation}

\subsection{eROSITA X-ray observations}
\label{sec:x-ray}

We use X-ray imaging of the A3391/95 field from SRG/eROSITA as presented in \citet{reiprich_2021} and \citet{bruggen_2021}. To obtain a map for on-target and off-target radio source classification, count and exposure images were constructed in the $0.2-2.3$ keV band for each observation and combined into mosaicked products over the full field.

To highlight diffuse emission while suppressing small-scale noise, we produced a wavelet-filtered X-ray image using \textsc{wvdecomp}\footnote{\url{https://github.com/avikhlinin/wvdecomp}}. The merged counts image was decomposed into wavelet scales and filtered by selecting structures on scales up to $\sim27\arcmin$, using a scale-dependent detection threshold that decreases from $5\sigma$ on the smallest scales to $2.8\sigma$ on the largest to control false positives. The filtered counts map was then corrected for spatial variations in sensitivity by dividing by the merged exposure map. On the largest scale, corresponding to $\sim54\arcmin$, we applied an additional Gaussian taper to suppress large-scale noise amplified by the exposure correction. Since this wavelet-filtered image is used only for visualization and to extract contours of the diffuse emission for on-target and off-target classification of polarized radio sources (see Sec.~\ref{sec:rm_grid}), this step does not affect any quantitative X-ray measurements. The resulting wavelet-filtered X-ray image is shown in Fig.~\ref{fig:x_ray}, with Stokes~$I$ radio emission overlaid.

\subsection{MeerKAT radio observations}
\label{sec:radio}

\begin{figure*}
    \centering
    \includegraphics[width=\textwidth]{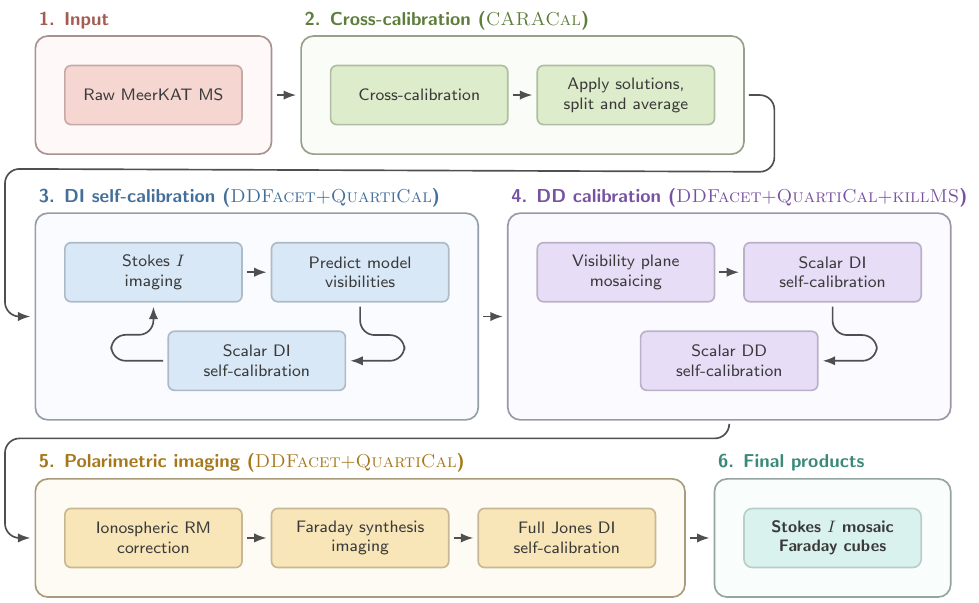}
    \caption{Schematic summary of the MeerKAT data reduction workflow.}
    \label{fig:pipeline}
\end{figure*}

The targets were observed over three epochs between December 4, 2022 and February 10, 2023 under proposal SCI-20220822-FG-01. Table~\ref{tab:observation} summarizes the target fields and calibrators. Each dataset consists of three pointings: a central field and two offsets located 25$\arcmin$ east and west of the center. The correlator delivered 4096 channels across 856 MHz centered at 1284 MHz, with a spectral resolution of 208.984 kHz and an integration time of 8 seconds. Each pointing was observed for 2 hours in each epoch, for a total of 18 hours on source across all observations. In Fig.~\ref{fig:pipeline} we provide a summary of the steps taken to reduce the MeerKAT data.

\begin{table*}[t]
    \centering
    \caption{Observation summary.}
    \label{tab:observation}
    \small
    \setlength{\tabcolsep}{6pt}
    \begin{tabular}{@{}l l l l l@{}}
    \doublehline
    \textbf{Name} & \textbf{Role} & \textbf{Use} & \textbf{RA} & \textbf{Dec} \\
    \hline
    \multicolumn{5}{@{}l@{}}{\textbf{Calibrators}} \\
    \hline
    J0408$-$6545 & Primary      & Flux scale, bandpass, delay & 04:08:20.380 & $-$65.45.09.100 \\
    J0538$-$4405 & Secondary    & Time-dependent gains        & 05:38:50.000 & $-$44.05.09.000 \\
    J0521$+$1638 & Polarization & Cross-hand phase            & 05:21:09.890 & $+$16.38.22.100 \\
    \hline
    \multicolumn{5}{@{}l@{}}{\textbf{Targets}} \\
    \hline
    offset-1 & Target & Science field                         & 06:24:19.800 & $-$54:04:58.000 \\
    bridges  & Target & Science field                         & 06:26:48.600 & $-$54:04:58.000 \\
    offset-2 & Target & Science field                         & 06:29:17.400 & $-$54:04:58.000 \\
    \hline
    \end{tabular}
\end{table*}

\subsection{Cross-calibration}

We performed cross-calibration with the CARACal pipeline \citep{caracal_2020}. We first split the calibrator scans and flagged auto-correlations, shadowed antennas, and channels affected by the H\textsc{i} 21-cm line. Radio-frequency interference (RFI) was removed using automated flagging with the \textsc{tricolour} package \citep{tricolour_2022}, which applies robust time–frequency thresholding and morphological filtering. Bandpass, instrumental delay, and the absolute flux-density scale were determined from J0408-6545 using standard \textsc{casa} tasks. Assuming this source is unpolarized, we solved for the instrumental polarization on the same calibrator. Time-dependent complex gains were derived from the secondary calibrator J0538-4405. The cross-hand phase was tied using the polarized calibrator J0521+1638, adopting its catalog polarization model \citep{perley_2013}.

Once we were satisfied with the solutions, the target fields were split from the measurement sets, while applying calibration solutions and automatic flagging for RFI using \textsc{tricolour}. Frequency channel averaging was performed with a factor of 4, lowering the resolution to 836 kHz.

\subsection{Scalar direction-independent self-calibration}

\begin{table}
    \centering
    \caption{Imaging setup used during self-calibration and final imaging.}
    \begin{tabular}{@{}p{0.4\columnwidth} p{0.4\columnwidth}@{}}
    \doublehline
    \multicolumn{2}{@{}l@{}}{\textbf{General}} \\
    \hline
    Field of view & $3\degree\times2\degree$ \\
    Pixel size & 2.0$\arcsec$ \\
    Weighting & Briggs (robust = 0) \\
    Frequency range & $890-1667$ MHz \\
    Frequency resolution & 7.8 MHz \\
    \hline
    \multicolumn{2}{@{}l@{}}{\textbf{Stokes $I$}} \\
    \hline
    Resolution & $5.7\arcsec \times 5.4\arcsec$ \\
    Polynomial order & 3 \\
    Weighting & Multifrequency synthesis (MFS) \\
    \hline
    \multicolumn{2}{@{}l@{}}{\textbf{Faraday synthesis}} \\
    \hline
    Resolution & $9\arcsec \times 9\arcsec \times 44.9$ \text{rad m$^{-2}$} \\
    Taper & $7\arcsec$ \\
    Maximum Faraday depth & $\pm 400$ rad m$^{-2}$ \\ 
    Faraday depth sampling & 5 rad m$^{-2}$ \\
    \hline
    \end{tabular}
    \tablefoot{The frequency range and frequency resolution listed here refer to the channels used internally by \textsc{ddfacet} during imaging.}
    \label{tab:imaging}
\end{table}

From the cross-calibrated target data, we built an initial Stokes $I$ sky model with \textsc{ddfacet} \citep{tasse_2018}, using the Högbom deconvolution mode. Imaging and deconvolution parameters are summarized in Table~\ref{tab:imaging}. Each field was tessellated into facets centered on the brightest sources using \texttt{MakeModel.py} from the \textsc{ddf} software \footnote{A MeerKAT direction-dependent imaging guide for \textsc{ddfacet} is available on the project wiki (\url{https://github.com/saopicc/DDFacet/wiki}).}. Deconvolution was first run for two major iterations using automasking with an aggressive minor-cycle threshold, then generated a mask from the apparent (primary beam attenuated) image using \texttt{MakeMask.py}. The thresholding was performed using a noise map, which was evaluated using sliding windows of 100 pixels. The deconvolution was then resumed from the residual visibilities until convergence. Model visibilities were generated from a CLEAN component dictionary containing component positions and polynomial coefficients for the frequency dependence. The coefficients were evaluated at the degridding frequencies to construct a frequency-dependent sky model. This model was then Fourier transformed to the visibility domain and degridded with the time- and frequency-dependent primary-beam models from \textsc{eidos} \citep{eidos_2021}, interpolated to a frequency resolution of 7.8 MHz.

We performed initial, per-epoch, per-pointing self-calibration  with \textsc{quartical} \citep{quartical_2023} using the constructed models in three rounds: two rounds of scalar phase calibration with 60 second solution intervals, followed by one round of scalar complex calibration with 30 minute solution intervals. All three rounds used the full band to derive the solutions. After each round we reimaged the field, and updated the sky model by predicting visibilities from the current component list. We used a CLEAN mask threshold of 8, 6 and 5$\sigma_I$ for the first, second and third round of imaging, respectively.

From the individually calibrated datasets, we performed joint deconvolution of all pointings using the visibility plane mosaicing mode of \textsc{ddfacet}. In this mode, the visibilities from each pointing are internally shifted to a common phase center and gridded onto a single uv-plane, which is Fourier–transformed to produce a common dirty image for deconvolution. This approach has recently demonstrated strong potential, achieving a 50\% higher dynamic range in the Shapley Supercluster Core \citep{Trehaven_2025}, due to the deeper deconvolution made possible by combining more data during the deconvolution process. We applied a $5\sigma$ clean mask to deconvolve the mosaic and predicted the resulting model components back into the corresponding measurement sets. We found that the obtained mosaic contained significantly more calibration artifacts than the individual pointing images. We attribute this to inconsistencies between the independently self-calibrated datasets, most likely caused by differences in the sky models used for each pointing. The resulting measurement sets are therefore not necessarily on a common amplitude and phase scale, which can make residual artifacts more prominent in the mosaic. We found that a single round of direction-independent calibration over the full time range at a single frequency, followed by solutions per time interval over the full bandwidth, significantly improved the image quality.

\subsection{Direction-dependent self-calibration}

Given the large field of view and the presence of bright, extended sources near the steepest part of the primary beam, direction-dependent calibration was required. We performed this calibration with the \textsc{killms} software \citep{smirnov_2015, killms_2023}, using the model component list derived from the mosaic. We used the non-linear Kalman Filter algorithm (KAFCA) to solve for scalar gains, using a solution interval of 5 minutes, in 32 frequency bins. We selected ten solution directions, centered on bright sources that were responsible for the dominant imaging artifacts, which can be seen in Fig.~\ref{fig:x_ray}. We found that this tessellation provided a good balance between keeping sufficient flux in each solution direction and limiting the number of directions to a manageable level for calibration. In addition, the tessellation ensured that no bright source extended across multiple facets, which reduced boundary artifacts and improved the stability of the direction-dependent solutions. We found that direction-dependent calibration was the most crucial for reducing artifacts around bright sources. Further experiments with the tessellation scheme, and increasing the flexibility of the model could potentially help improve the quality further. However, for our science case, these results were deemed sufficient.

\subsection{Full Jones direction-independent self-calibration}
\label{sec:full_jones}

As a final calibration step, we aimed to remove any non-scalar effects remaining in the data. We first corrected for the time-dependent ionospheric Faraday rotation using \textsc{spinifex} \citep{spinifex_2025}, which estimates the total electron content (TEC) and thus the ionospheric RM from satellite-based TEC data. Using the metadata in the measurement set, \textsc{spinifex} derives the time-dependent ionospheric RM at the pointing center of the observation. Recent work by \cite{perley_2026} has shown that this approach can overestimate the ionospheric RM correction. However, for MeerKAT, this overcorrection is minimal ($\sim 0.3$ rad m$^{-2}$) and is not expected to affect our results significantly. We then used \textsc{dp3} \citep{DP3_2018} to apply the \textsc{spinifex} RM estimates directly to the measurement sets.

Next we imaged the data in full polarization, using the Faraday synthesis deconvolution mode of \textsc{ddfacet}. The full algorithm is described in \cite{gustafsson_2025}, but here we list the elements specific to this work:

\begin{enumerate}
\item Due to the wide field of view and the strong frequency dependence of the primary beam, we did not apply a primary beam correction to the residual cube during deconvolution. Instead, during deconvolution we use the primary-beam-corrupted,
direction-dependent three-dimensional $(l,m,\nu)$ point spread
function (PSF). This introduces a frequency-dependent slope in the PSF in facets far from the pointing center, resulting in a broader main lobe in $\phi$ of the $(l,m,\phi)$ PSF. This approach has two advantages. First, it produces a noise level in the residual Faraday cube that is flatter across the field, which leads to more stable deconvolution. Second, it allows us to use direction-independent spectral weights, which simplifies the imaging steps and avoids position-dependent weighting effects. After deconvolution, we convolve the clean components with a 3D restoring beam, and add them to the final primary-beam-corrected residual Faraday cube. 

Because the primary-beam correction makes the noise both position and frequency dependent, we applied spatially varying channel weights when forming the final residual Faraday cube. This was done to improve the sensitivity and reliability of the RM measurements. The direction-independent channel weights were scaled at each pixel by the local primary-beam attenuation, thereby down-weighting channels in which the primary-beam correction strongly amplifies the noise. This is motivated by the weighted RM synthesis formalism of \cite{brentjens_2005}, who noted that primary-beam attenuation leads to a position-dependent weight function and hence a local RM spread function. This weighting scheme increased the average polarized-source density by about 50\%, with the strongest improvement toward the edges of the field.

\item To complement the less restrictive external Stokes $I$ mask, a three-dimensional automask is generated at the beginning of each major cycle. The automask is based on noise maps produced from the bandwidth-averaged Stokes $Q$ and $U$ residual maps. We set a threshold of $5 \sigma_{QU}$ which removes the majority of unpolarized sources, and speeds up deconvolution significantly as peak finding is only performed on a small fraction of the voxels in the Faraday cube.
\end{enumerate}

While we do not deconvolve Stokes $V$, we still grid and Fourier transform it in order to apply the image plane correction described by \citet{tasse_2018}. This correction applies a $4\times4$ Mueller-matrix operation in the image domain to account for instrumental polarization mixing. Its purpose is to make the effective PSF and beam response behave approximately like a scalar convolution within each frequency chunk, so that the imaging response is more uniform between Stokes parameters across the field. Because we do not expect significant astrophysical Stokes $V$ emission in this field, we do not include Stokes $V$ components in the sky model. When predicting model visibilities for the final calibration step, we therefore use only Stokes $IQU$ CLEAN components. We then run \textsc{quartical}, solving for full $2\times2$ complex gains in frequency intervals of 64 channels and eight integrations in time. This term captures residual amplitude and phase errors as well as polarization leakage, which helped reduce residual artifacts around the brightest polarized sources.

\section{Final imaging}
\label{sec:final_imaging}

In this section we go over the final imaging steps of our pipeline, and briefly discuss the quality of our Stokes $I$ and polarized images.

\subsection{Total intensity imaging}

\begin{figure*}
    \centering
    \includegraphics[width=\textwidth]{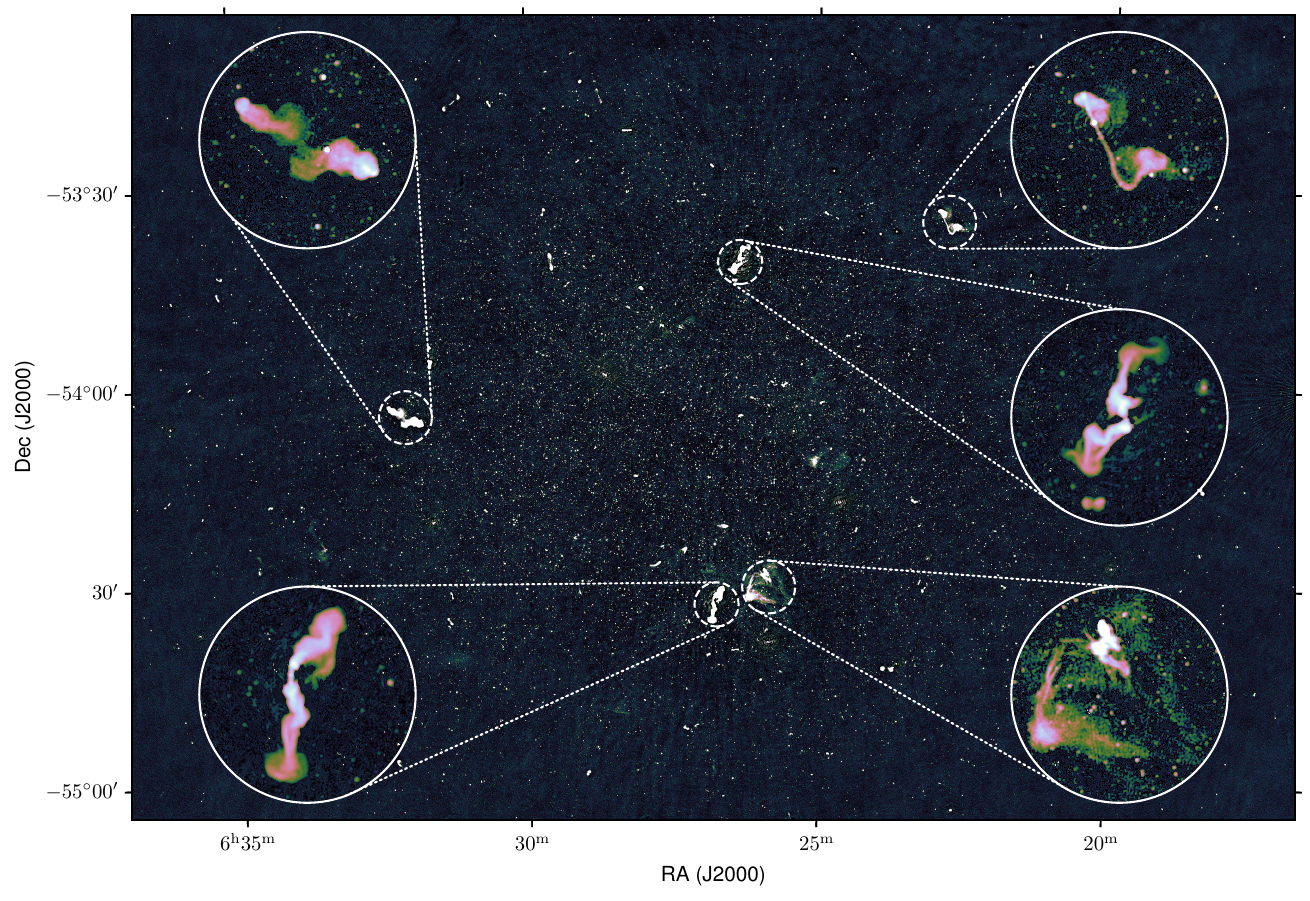}
    \caption{Stokes I MFS ($890-1667$ MHz) mosaic of the A3391/95 system and surrounding sources. To highlight the number of unresolved sources, the background is in linear scale and saturated at 25 $\mu$Jy beam$^{-1}$. The insets show interesting extended radio sources, in logarithmic scale to highlight internal structure. The image is at a resolution of $5.7\arcsec \times 5.4\arcsec$ and has a noise level of 2.5 $\mu$Jy beam$^{-1}$.}
    \label{fig:StokesI}
\end{figure*}

The final high-resolution Stokes $I$ mosaic was imaged using MFS to improve the quality of the frequency-integrated image. No uv-taper was applied, since the image was used as a deep continuum reference and not for spectral analysis. The Stokes $I$ image can be seen in Fig.~\ref{fig:StokesI}, at a resolution of $5.7\arcsec \times 5.4\arcsec$. We estimate the rms noise from the median absolute deviation (MAD)\footnote{We use the MAD throughout this work because it is less sensitive to outliers than the standard deviation.} of the negative pixels within the 50\% beam power region, mirroring the distribution to reconstruct the full Gaussian noise, which yields an rms of 2.5 $\mu$Jy beam$^{-1}$. From this noise estimate we computed the dynamic range as the peak flux density divided by the rms, yielding a dynamic range of $8.6\times10^4$. Due to the extensive calibration, the resulting image is almost free of artifacts, with the exception of some bright compact sources. The source density in the image is staggering, with a clear reduction toward the edges, where the telescope sensitivity decreases rapidly. Some interesting extended sources in the field which have already been reported in \cite{bruggen_2021} are highlighted. The polarized radio properties of these sources are discussed briefly in Appendix~\ref{sec:appendix}. A few regions of diffuse emission are detected, but no large-scale structure is observed at this resolution. This non-detection is unlikely to be caused by missing short baselines. The shortest unflagged baseline in our datasets is $b_{\rm min} \simeq 29$ m, corresponding to a largest recoverable angular scale of $\theta_{\rm LAS} \simeq 0.6\lambda/b_{\rm min} \simeq 24\arcmin$ at the lower end of the band. This is similar to the angular scale adopted for the bridge region in this work (Fig.~\ref{fig:x_ray}). We therefore expect the data to be sensitive to bridge emission with structure on smaller angular scales, although a very smooth component on scales larger than $\sim24\arcmin$ could be partly resolved out. We therefore attempted point-source subtraction followed by low-resolution imaging to improve sensitivity to diffuse emission on larger angular scales. This procedure did not reveal any additional diffuse emission across the field. The low resolution images produced are however used in Sec.~\ref{sec:upper_limit} to quantify upper limits of the radio bridge emissivity.

\subsection{Polarization imaging}

For the final full-polarization imaging, a uniform angular resolution across the band was required, since frequency-dependent resolution can introduce artificial complexity in the residual Faraday spectra if left uncorrected. We therefore applied a uv-taper by multiplying the visibilities with a Gaussian, corresponding to a full width at half maximum (FWHM) of $7\arcsec$ in the image plane. This yielded an approximately uniform synthesized beam of $9\arcsec$ across the band. Stokes $Q$ and $U$ imaging was performed using Faraday synthesis, with the imaging parameters shown in Table~\ref{tab:imaging}, as discussed in Sec.~\ref{sec:full_jones}. With the selected frequency resolution, we estimate the maximum detectable Faraday depth $||\phi_{\text{max}}||$, following \citep{brentjens_2005}

\begin{equation}
    ||\phi_{\text{max}}|| = \frac{\sqrt{3}}{\delta\lambda^2} = 2143 \text{rad m}^{-2},
\end{equation}

where $\delta\lambda^2$ is the average spacing in $\lambda^2$. Hence, the maximum observable depth is much higher than the maximum Faraday depth in our produced Faraday cubes, and no significant bandwidth depolarization is expected at this frequency resolution.

Using the same noise estimator as for Stokes $I$, we find rms levels of $6.0$, $3.2$, $3.5$, and $3.1~\mu$Jy~beam$^{-1}$ in Stokes~$I$, $Q$, $U$, and $V$, respectively, at this lower resolution. 

\begin{figure*}
    \centering
    \includegraphics[width=\textwidth]{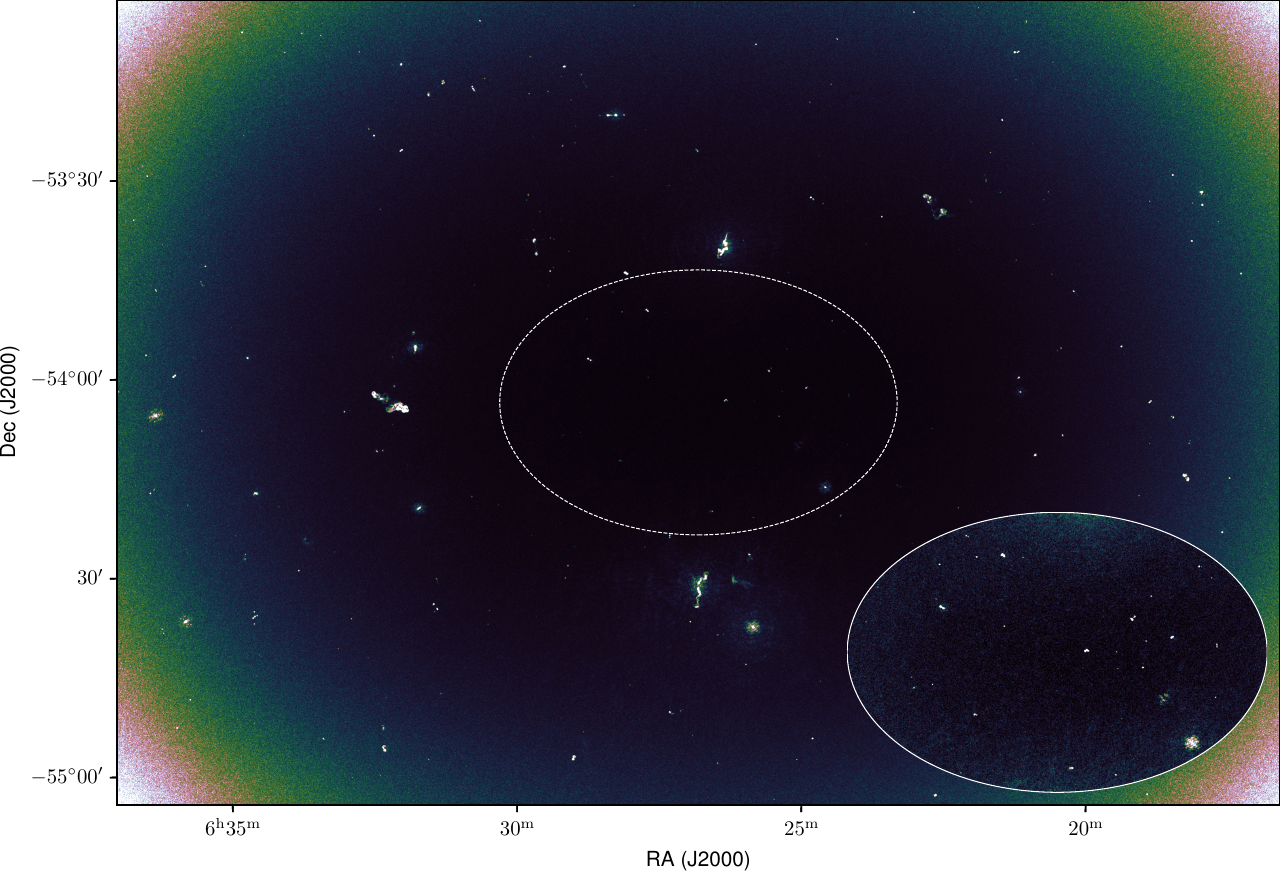}
    \caption{Rician debiased peak polarized intensity map of the A3391/95 system and surrounding sources. The image is at a resolution of $9\arcsec \times 9\arcsec$. The bright corners are due to the correction of the primary beam attenuation. The inset shows the central field at a higher contrast, to highlight the polarized source density which is not visible in the full image.}
    \label{fig:StokesP}
\end{figure*}

In Fig.~\ref{fig:StokesP} we show the peak polarized intensity mosaic, corrected for Rician bias following \cite{George_2012}. The bias is estimated from a noise map, computed from the wings ($|\phi|>200$ rad m$^{-2}$) of the Faraday cube. Most sources in the field are compact, with the exception of the bright radio jets. The image still contains artifacts around the brightest polarized sources, which we attribute to direction-dependent effects that are not captured by the scalar calibration used in the current \textsc{killms} setup. A full-Jones direction-dependent solution could, in principle, reduce these artifacts, but it also introduces many additional degrees of freedom. These extra parameters may not be sufficiently constrained by the available sky model, which could in turn lead to unstable calibration solutions and potentially degrade the final image quality. Furthermore, the negative lobes, typically seen around sharp edges in total intensity images appear as, due to the non-negativity of polarized images, artificial extension around these sources.

\section{Rotation measure grid}
\label{sec:rm_grid}

To construct the catalog of polarized sources and the corresponding RM grid, we followed the procedure given below:

\begin{enumerate}
    \item We constructed a polarized noise map from the wings of the Faraday spectrum at $|\phi| > 200$ rad m$^{-2}$, where no significant astrophysical signal is expected. At each pixel, the noise was estimated from the MAD of the real and imaginary parts of the Faraday spectrum separately. The final polarized noise map, $\sigma_{QU}$, was defined as the mean of these two maps. The Stokes $I$ noise map was derived from the bandwidth-averaged restored image by measuring the rms in sliding windows of 100 pixels.
    \item We constructed a polarization detection mask by selecting pixels with peak polarized intensity above $5\sigma_{QU}$. We then applied an additional $5\sigma$ Stokes $I$ mask to reduce spurious polarized detections due to the positively biased Rician noise distribution in the polarized intensity.
    \item We ran \textsc{SoFiA} \citep{serra_2015} on the combined mask to group connected detections and produce the final catalog of polarized sources. Detections separated by less than two beam widths were merged into a single source.
    \item For each source identified by \textsc{SoFiA}, we fitted a 1D Gaussian to the Faraday spectrum of every pixel within the source mask. A single RM value for each source was then obtained by computing the polarized intensity weighted mean of all fitted pixel RMs.
\end{enumerate}

With this procedure, we identified a total of 434 polarized sources within the field, corresponding to a source density of 73 sources per square degree. The polarized-source density varies strongly across the mosaic because of the changing sensitivity. Since our wide-field imaging extends well beyond the half-power point of the primary beam, the field-averaged density is reduced by the outer regions, where the source density is about 30 sources per square degree. In the central part of the field, near the A3391/95 system and at higher sensitivity, the source density rises to about 110 sources per square degree. This is broadly consistent with expectations for SKA1-MID surveys, which predict polarized-source densities of roughly 60 to 90 sources per square degree at similar sensitivities \citep{Heald_2020}.

\begin{figure}
    \centering
    \includegraphics[width=\columnwidth]{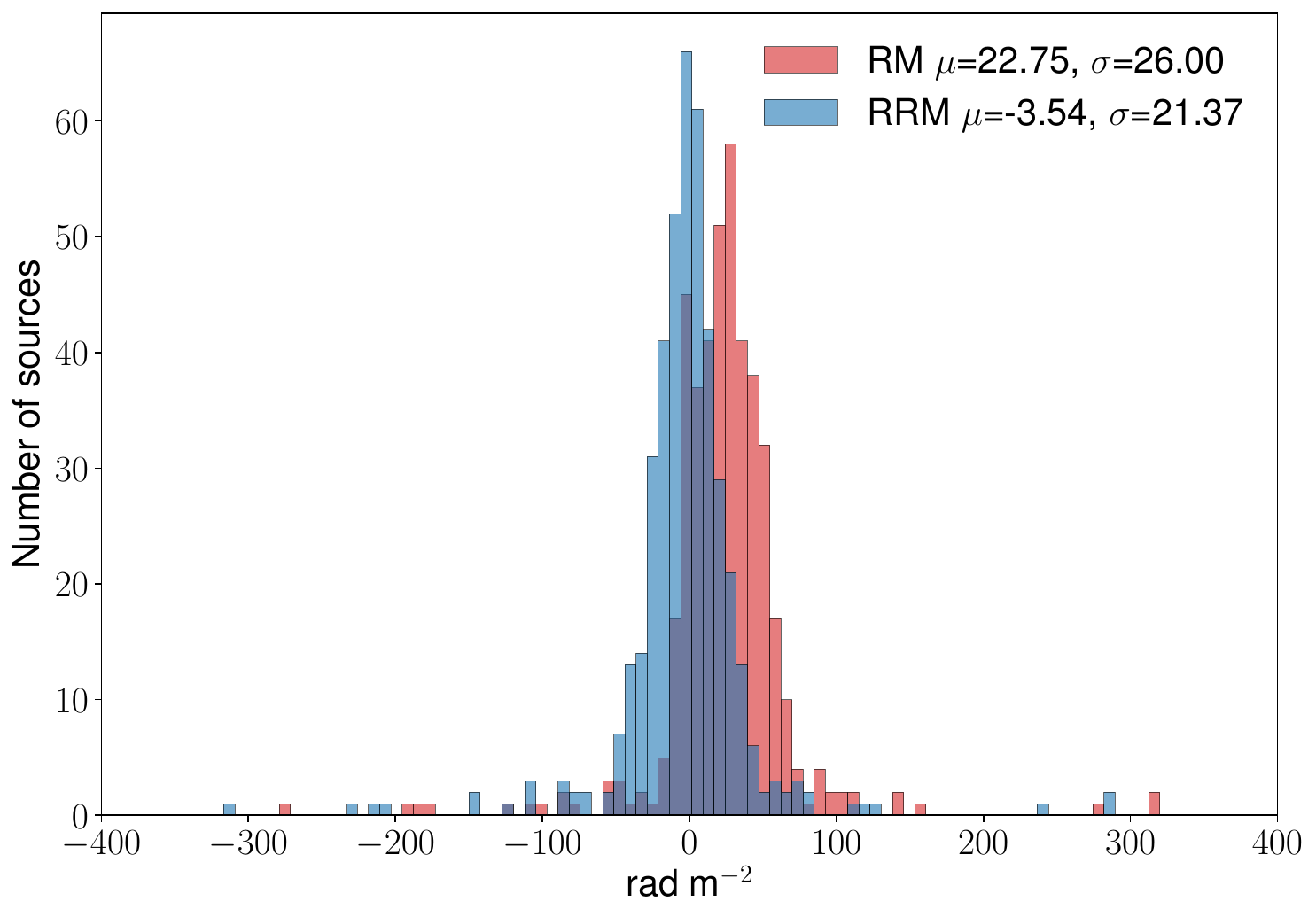}
    \caption{Distribution of RMs before (red) and after (blue) subtraction of the GRM, together with their reported median and standard deviation.}
    \label{fig:rm_histogram}
\end{figure}

\begin{figure}
    \centering
    \includegraphics[width=\columnwidth]{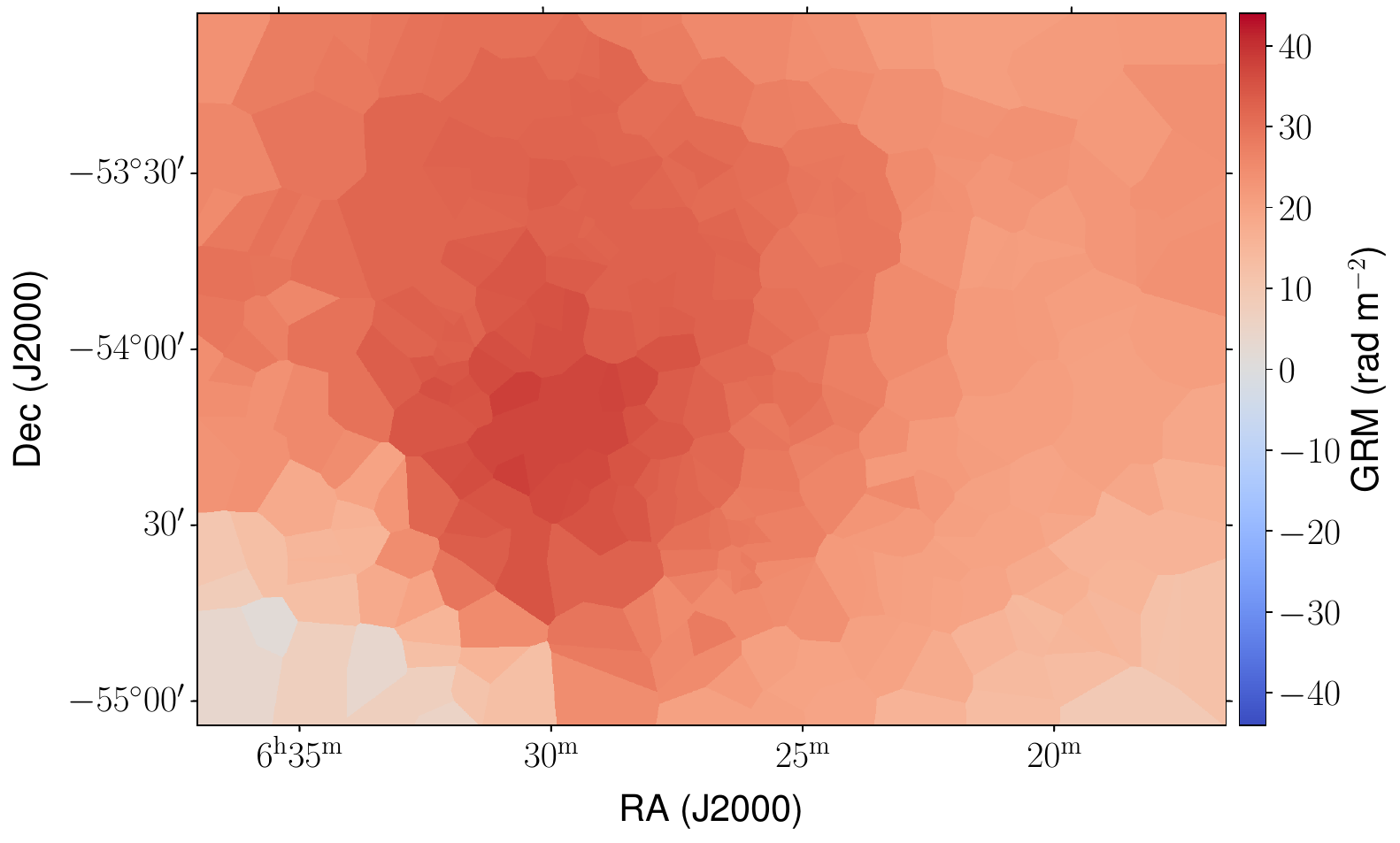}
    \caption{GRM map obtained using the median annulus method. The Galactic contribution is smooth on degree scales, but with a sharp jump in the southeast.}
    \label{fig:RM_gal}
\end{figure}

Galactic rotation measure (GRM) was removed using the median annulus method \citep{anderson_2024}. The inner radius was set to twice the angular size of the largest source in the field to avoid self-contamination of the GRM estimate, corresponding to 0.1 deg. Because the density of polarized sources varies across the field, we adopted a fixed outer radius of 0.8 deg to ensure a consistent angular scale for the GRM estimation. When instead setting the outer radius to include a fixed number of sources, as in \citet{anderson_2024}, the effective angular scale changes with position. In dense regions the annulus probes small-scale fluctuations, while in sparse regions it averages over much larger scales. This led to GRM estimates that either followed fine-scale structure unlikely to be Galactic, or over-smoothed the large-scale GRM gradient. With a fixed outer radius, the number of sources per annulus typically ranges from 67 (10th percentile) to 190 (90th percentile), with a median of 131 (50th percentile). In Fig.~\ref{fig:RM_gal} we show the GRM map of the observed field. We note a broadly smooth GRM gradient across the field, as expected for a large-scale Galactic foreground, together with a localized sharp transition in the southeast. Fig.~\ref{fig:rm_histogram} shows the distribution of RMs before and after subtraction of the GRM estimate. The residual rotation measure (RRM) distribution has a median close to zero, together with a lower scatter compared to the RM, indicating that the GRM has successfully been subtracted.

To further assess whether the RRM is contaminated by the Milky Way interstellar medium, we tested for correlations between the RM measurements and tracers of Galactic gas using the Spearman rank correlation coefficient $\rho_{\rm S}$. We used the H\textsc{i} column-density map \cite{HI4PI_2016} and the H$\alpha$ intensity map \cite{Haffner_2010}. For the full sample, the total $|{\rm RM}|$ shows weak positive correlations with both H$\alpha$ and H\textsc{i}, with $\rho_{\rm S}=0.151$ (p-value$=0.0016$) and $\rho_{\rm S}=0.113$ (p-value$=0.018$), respectively. In contrast, $|{\rm GRM}|$ is strongly correlated with H$\alpha$, with $\rho_{\rm S}=0.465$ (p-value$=1.2\times10^{-24}$), but shows no significant correlation with H\textsc{i}, with $\rho_{\rm S}=-0.075$ (p-value$=0.12$). The corrected $|{\rm RRM}|$ shows only a weak anticorrelation with H$\alpha$, with $\rho_{\rm S}=-0.136$ (p-value$=0.0045$), but remains significantly correlated with H\textsc{i}, with $\rho_{\rm S}=0.303$ (p-value$=1.1\times10^{-10}$).

To test whether these residual correlations are driven by sources near the mosaic boundaries, we repeated the analysis after restricting the sample to progressively more central regions of the field, defined by the distance to the nearest image edge. The correlation between $|{\rm RRM}|$ and H\textsc{i} remains stable, with $\rho_{\rm S}\approx0.32$--0.34 for all central subsamples, while the strong correlation between $|{\rm GRM}|$ and H$\alpha$ is also preserved. We therefore find no evidence that the residual $|{\rm RRM}|$-H\textsc{i} correlation is primarily caused by poorer corrections at the edges of the field. Instead, the GRM correction appears to remove the dominant large-scale ionized Galactic foreground traced by H$\alpha$, while a residual component correlated with H\textsc{i} remains.

\begin{figure}
    \centering
    \includegraphics[width=\columnwidth]{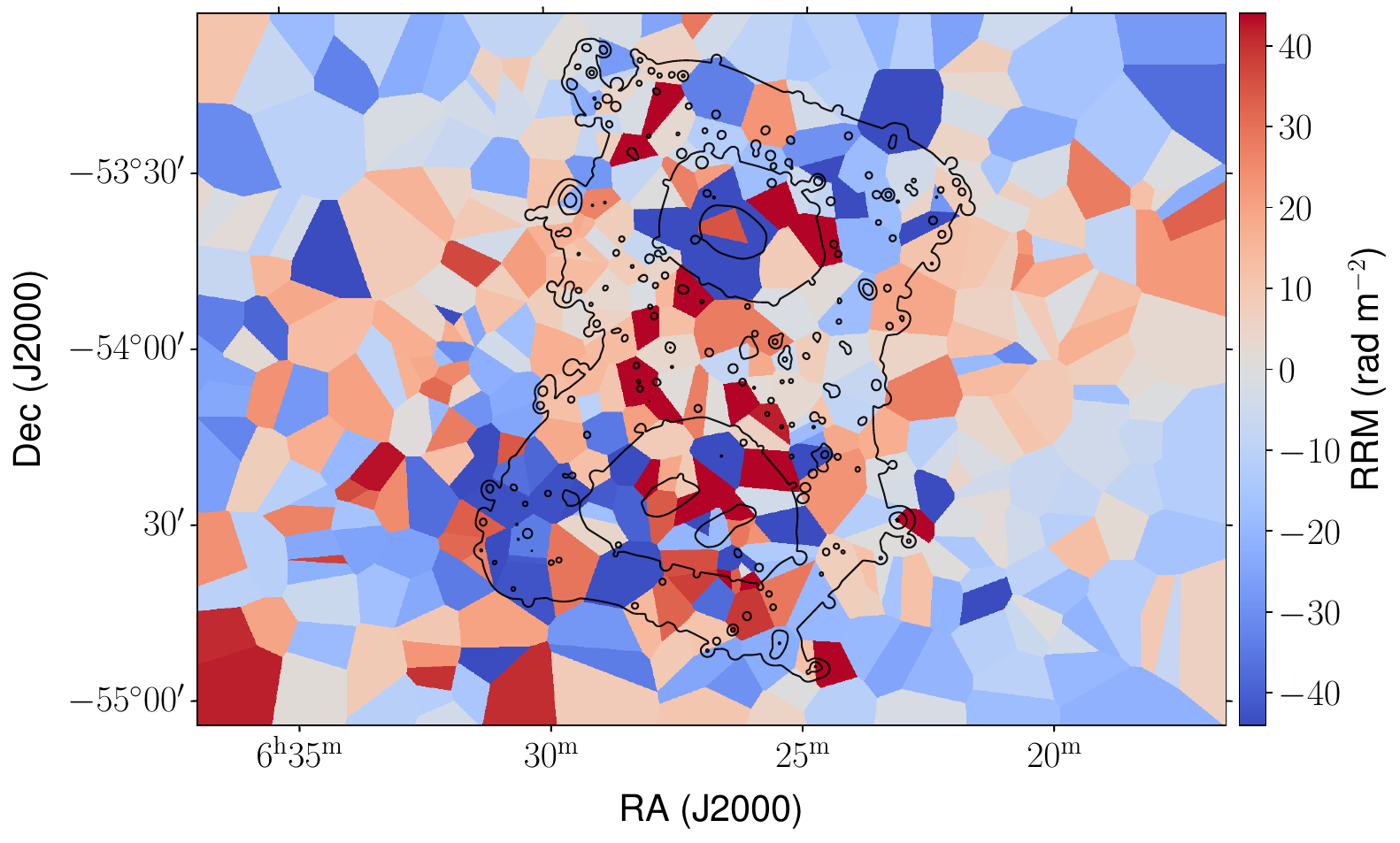}
    \caption{Residual RM map of the A3391/95 field, generated by nearest neighbor interpolation of the discrete RM measurements. For visualization purposes, the colorbar is clipped at the 90th percentile. Black contours trace the diffuse X-ray surface brightness (ICM/bridge emission) from the exposure-corrected, wavelet-filtered eROSITA map.}
    \label{fig:RM_tessel}
\end{figure}

In Fig.~\ref{fig:RM_tessel} we show our RRM grid of the field, as a nearest-neighbor interpolation from the RRM catalog. In contrast to the large gradient of the GRM, the RRM shows an irregular pattern together with a large scatter of values. The sensitivity of our observation is clearly visible, as the central region has a much higher density of sources.

Hereafter, unless stated otherwise, we denote the RRM simply as RM, except where needed to avoid confusion. For the analyses presented below, we also tested whether possible embedded cluster members affect the results by excluding sources with known redshifts close to the mean cluster redshift of the system. This did not produce any significant change in the derived source statistics. As a first inspection of the RM distribution of our field, we classified the polarized sources as on/off-target, determined by the lowest contour of X-ray emission, shown in Fig.~\ref{fig:RM_tessel}. We quantified the RM fluctuations by comparing the scatter of the on-target and off-target RM distributions. The excess RM scatter was estimated following an approach similar to \citet{lopez_2025}, with the measurement uncertainties and the uncertainty of the local GRM correction propagated through Monte-Carlo realizations. The full procedure is described in Appendix~\ref{app:rm_scatter_depol}.

The resulting RM scatters and excess scatter are listed in Table~\ref{tab:rm_scatter}. We find an on-target RM scatter of $29.31 \pm 1.26$ rad m$^{-2}$ and an off-target scatter of $18.86 \pm 0.69$ rad m$^{-2}$, corresponding to an excess RM scatter of $22.42 \pm 1.73$ rad m$^{-2}$. This excess is detected at a statistical significance of 13.0$\sigma$. The off-target scatter is higher than expected for a random extragalactic background \citep{schnitzeler_2010}. However, at the Galactic latitude of A3391/95 ($b \simeq -25^\circ$), the observed RM scatter is expected to be dominated by the Galactic foreground and to increase toward lower Galactic latitude approximately as $1/\sin |b|$ \citep{schnitzeler_2010}. Following the model in \cite{schnitzeler_2010} for the expected RM scatter

\begin{equation}
    \sigma_{\text{RM}} = \sqrt{\big(\frac{\sigma_{\text{RM,MW}}}{\sin{|b|}}\big)^2 + \sigma_{\text{RM,EG}}^2},
\end{equation}

where $\sigma_{\text{RM,MW}}$ and $\sigma_{\text{RM,EG}}$ are the Galactic and extragalactic contributions to the observed RM scatter, respectively. Using the best fit parameters for negative Galactic latitudes ($\sigma_{\text{RM,MW}}=8.4 \pm 0.1$ rad m$^{-2}$, $\sigma_{\text{RM,EG}}=5.9 \pm 0.2$ rad m$^{-2}$), the expected RM scatter at $|b| = 25.23^\circ$ is $\sigma_{\text{RM}} = 20.57 \pm 0.23$ rad m$^{-2}$. This is broadly consistent with the measured off-target scatter in our field, considering any Galactic foreground fluctuations and small-scale structure along the line of sight at this latitude. It should be noted that some of the GRM scatter has been removed when subtracting the smooth gradient seen in Fig.~\ref{fig:RM_gal}. Without this subtraction, the RM scatter of the off-target region is slightly higher at $21.76 \pm 0.61$ rad m$^{-2}$, and therefore more consistent with \cite{schnitzeler_2010}.

\begin{table}[h]
    \centering
    \caption{RM scatter measurements for the on-target and off-target samples, and the derived excess scatter.}
    \begin{tabular}{@{}p{0.34\columnwidth} p{0.28\columnwidth} p{0.28\columnwidth}@{}}
    \doublehline
    \textbf{Sample} & \textbf{N} & \textbf{$\sigma_{\text{RM}}$ (rad m$^{-2}$)} \\
    \hline
    on-target  & 171 & $29.31 \pm 1.26$ \\
    off-target & 263 & $18.86 \pm 0.69$ \\
    \hline
    excess     & --- & $22.42 \pm 1.73$ \\
    \hline
    \end{tabular}
    \tablefoot{Here, \textbf{N} is the number of sources in each sample.}
    \label{tab:rm_scatter}
\end{table}

To further divide the field, we defined four regions of interest, shown in Fig.~\ref{fig:x_ray}: the A3391 cluster, the A3395 system (including both the northern and southern components), the ESO-161 group, and the intercluster bridge region. The cluster and group regions were defined by their $r_{500}$ and centered on their X-ray peaks, whereas the bridge region was chosen to include the diffuse thermal emission connecting the clusters. We classified each polarized source as belonging to one or more of these regions based on its sky position. Since the regions are not mutually exclusive, a given source can be assigned to multiple regions. We performed the same Monte-Carlo scatter estimation as before, but now computed the RM scatter separately for each of the four regions of interest. Sources falling within the lowest X-ray contour but outside all four regions were excluded. The resulting scatters are presented in Table~\ref{tab:rm_scatter_regions}. We found that while the clusters show a significant RM scatter excess with respect to the off-target region ($\sigma_{\text{RM,A3391}}^{\text{excess}}=30.52 \pm 5.33$ rad m $^{-2}$, $\sigma_{\text{RM,A3395}}^{\text{excess}}=35.31 \pm 2.60$ rad m $^{-2}$), we found a smaller scatter in the bridge. For ESO-161, we observe a slight excess RM scatter of $11.36 \pm 5.00$ rad m $^{-2}$, however, only at a statistical significance of 2.27$\sigma$.

\begin{table}[h]
    \centering
    \caption{RM scatter measurements for each subregion and the off-target sample.}
    \begin{tabular}{@{}p{0.34\columnwidth} p{0.28\columnwidth} p{0.28\columnwidth}@{}}
    \doublehline
    \textbf{Sample} & \textbf{N} & \textbf{$\sigma_{\text{RM}}$ (rad m$^{-2}$)} \\
    \hline
    off-target & 263 & $18.86 \pm 0.69$ \\
    \hline
    A3391                        & 25  & $35.90 \pm 4.49$ \\
    A3395 (N/S)                  & 60  & $39.31 \pm 2.03$ \\
    ESO-161                      & 8   & $22.03 \pm 2.50$ \\
    bridge                       & 32  & $13.98 \pm 1.64$ \\
    \hline
    \end{tabular}
    \tablefoot{The scatter is reported as the median and $1\sigma$ uncertainty from the Monte-Carlo realizations. Here, \textbf{N} is the number of sources in each sample.}
    \label{tab:rm_scatter_regions}
\end{table}

\begin{figure}
    \centering
    \includegraphics[width=\columnwidth]{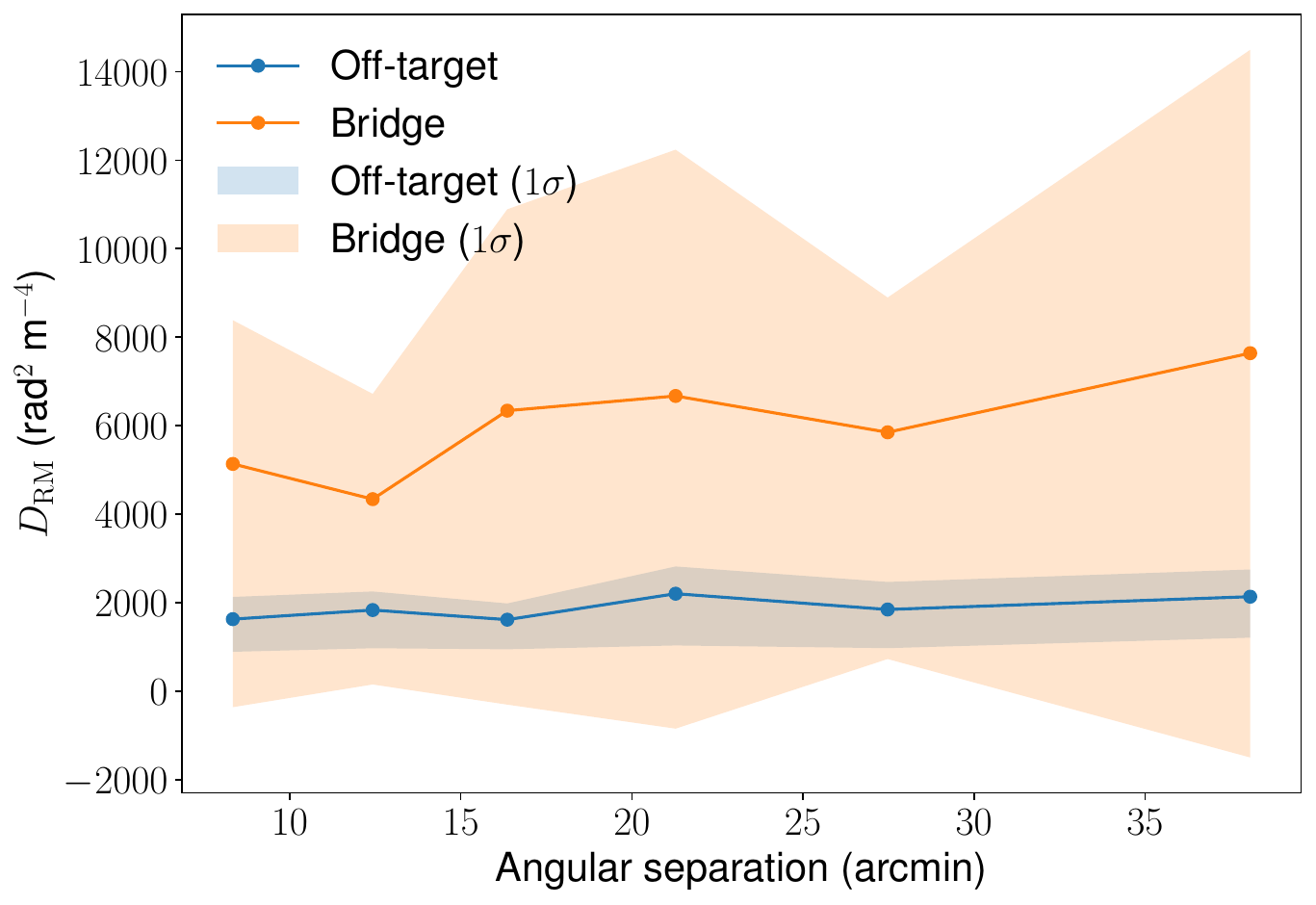}
    \caption{RM structure function for the bridge (orange) and the off-target region (blue). Points show the mean squared RM difference of source pairs binned by angular separation. Shaded bands indicate $1\sigma$ uncertainties from source bootstrap resampling.}
    \label{fig:structure-function}
\end{figure}

To test whether the comparatively low RM scatter measured in the bridge is simply a consequence of sampling a much smaller angular area than in the off-target region, we complemented the one-point RM-scatter analysis with a two-point statistic. We used the RM structure function \citep[e.g.,][]{minter_1996} to compare RM fluctuations in the bridge and the off-target region on matched angular scales. The structure function $D_{\rm RM}$ is defined as

\begin{equation}
    D_{\rm RM}(\delta\theta_k) = \frac{1}{N} \sum_{i=1}^N \big(\text{RM}(\theta) - \text{RM}(\theta + \delta\theta_k)\big)^2,
\end{equation}

where $\delta\theta_k$ are bins in angular separation between pairs of sources. We selected the bins to contain 75 pairs of sources per bin in the bridge, which, due to the higher number of off-target sources, yields a larger amount of source pairs. Furthermore, we excluded source pairs with an angular distance below 0.1 degrees, to remove any self-contamination from sources fragmented by the source mask. The resulting structure function is shown in Fig.~\ref{fig:structure-function}. The shaded $1\sigma$ ranges were obtained by bootstrap resampling of sources within each region followed by computing the MAD of the resulting distribution of $D_{\rm RM}$. The relatively large values of $D_{\rm RM}$ are caused by the squared pairwise RM differences and by a small number of high-$|{\rm RM}|$ outliers, which are visible in the full RM distribution shown in Fig.~\ref{fig:rm_histogram}. The bridge shows systematically higher $D_{\rm RM}$ values than the off-target sample, but the difference is not statistically significant. A permutation test under the null hypothesis that the bridge and off-target samples produce the same $D_{\rm RM}(\delta\theta)$ curve yields a one-sided p-value of 0.11. It should be noted that the bridge sample includes sources that also fall inside the group and cluster regions. If we remove those overlapping sources, too few sources remain in the bridge to fill the separation bins, and the structure function becomes too noisy to interpret.

To investigate the Faraday complexity of the polarized source population across the field, we first attempted to characterize each source using the second moment of the cleaned Faraday components and to compare this width to the RM spread function FWHM. However, for Faraday spectra containing multiple components or strongly non-Gaussian structure, we found that a single Gaussian fit did not robustly capture the complexity of the source. We therefore instead quantified the frequency-dependent depolarization of each source using the depolarization ratio (DP)

\begin{equation}
    \mathrm{DP} = \frac{p_{\rm lo}}{p_{\rm hi}},
\end{equation}
where $p_{\rm lo}$ and $p_{\rm hi}$ are the fractional polarizations in the low- and high-frequency subbands, respectively. The full procedure used to measure DP and propagate its uncertainty is described in Appendix~\ref{app:rm_scatter_depol}.

The resulting median depolarization ratio and corresponding errors for each target region are listed in Table~\ref{tab:depol_regions}. While A3391 and ESO-161 show clear signs of depolarization, A3395 is depolarized at a level comparable to that of the off-target sample. Furthermore, sources seen through the bridge region appear to experience similar or less depolarization than the off-target sources. This suggests that the magneto-ionic medium in the bridge region does not produce stronger depolarization on scales on the order of 10 kpc, comparable to our beam size at the mean redshift of the bridge system.

\begin{table}[h]
    \centering
    \caption{Median depolarization ratio for each subregion and the off-target sample.}
    \begin{tabular}{@{}p{0.34\columnwidth} p{0.28\columnwidth} p{0.28\columnwidth}@{}}
    \doublehline
    \textbf{Sample} & \textbf{N} & \textbf{DP} \\
    \hline
    off-target & 252 & $0.74 \pm 0.02$ \\
    \hline
    A3391                        & 24  & $0.59 \pm 0.02$ \\
    A3395 (N/S)                  & 58  & $0.75 \pm 0.03$ \\
    ESO-161                      & 8   & $0.66 \pm 0.06$ \\
    bridge                       & 32  & $0.78 \pm 0.02$ \\
    \hline
    \end{tabular}
    \tablefoot{Here, \textbf{N} is the number of sources in each sample.}
    \label{tab:depol_regions}
\end{table}

\section{Discussion}
\label{sec:discussion}

A key aspect of this work is the imaging and calibration strategy required to reach high-quality, wide-field polarimetric products at the depth needed for a dense and reliable RM grid. We combine direction-dependent calibration with visibility plane mosaicing, and we reconstruct deconvolved Faraday cubes using Faraday synthesis. Our workflow builds directly on the direction-dependent Faraday synthesis framework implemented in \textsc{ddfacet} \citep{gustafsson_2025} and on recent developments in direction-dependent visibility plane mosaicing with \textsc{ddfacet} and \textsc{killms} \citep{Trehaven_2025}. In the present work, we extend these ideas to full-Stokes imaging and Faraday-depth reconstruction in a target field, enabling deep deconvolution, while accounting for non-scalar effects modeled by the primary beam models, including polarization leakage, that would otherwise corrupt and bias full-Stokes polarimetric measurements.

The dense RM grid derived in Sec.~\ref{sec:rm_grid} enables a statistical study of Faraday rotation across A3391/95 and its surroundings. Using eROSITA X-ray observations to define on-target and off-target regions, we find a significant enhancement of the RM scatter in the on-target sample relative to the off-target control region. This result is primarily driven by the cluster regions, which show a substantially larger RM scatter compared to the off-target region. A similar increase in RM scatter toward cluster environments was recently reported by \cite{lopez_2025} for the Shapley Supercluster Core (SSC), using an RM grid constructed from ASKAP-POSSUM polarization data. This result is expected for polarized background sources seen through the magnetized ICM, where turbulent magnetic fields and spatial variations in the thermal electron density lead to enhanced Faraday-rotation fluctuations. \cite{lopez_2025} also reported an elevated RM scatter in the intercluster region of the SSC. In their analysis, the bridge region was defined to include the galaxy groups located between the two clusters, and they interpreted the resulting bridge excess as evidence that the supercluster RM signal is not produced only by the cluster interiors. In particular, they argued that the bridge contributes a significant fraction of the RM scatter excess measured over the full SSC field, and that the bridge scatter is comparable, within uncertainties, to the cluster values.

In contrast, the intercluster bridge in A3391/95 shows a comparatively low RM scatter of $13.98 \pm 1.64$ rad m$^{-2}$, which is below the off-target scatter of $18.85 \pm 0.69$ rad m$^{-2}$. Within the current uncertainties, this implies a nondetection of an additional contribution to the RM scatter above the off-target level. Since the bridge covers only a small angular area of the field, the RM scatter mainly reflects fluctuations on relatively small angular scales, whereas the off-target sample spans several square degrees. When we compare the bridge and off-target regions on the same angular scales using the RM structure function, we instead find a hint that the bridge has larger RM differences than the off-target region, although the effect is not statistically significant within the current uncertainties.

The depolarization ratio provides an additional probe of the magneto-ionic conditions in the A3391/95 field. In contrast to the RM scatter and the RM structure function, which trace changes in Faraday rotation between different background sources on scales from about 0.1 Mpc to several Mpc, depolarization reflects Faraday complexity within individual sources on orders of 10 kpc. In the system, A3391 shows clear evidence for enhanced Faraday complexity, consistent with the elevated RM scatter measured toward the cluster regions. This agreement supports the interpretation that the high cluster RM scatter is driven by magnetized and turbulent intracluster plasma rather than by large-scale foreground gradients alone. A3395 also shows enhanced RM scatter, but its depolarization ratio is comparable to that of the off-target sample. This suggests that the Faraday-rotating medium toward A3395 produces variations between different lines of sight, without introducing strong unresolved Faraday structure within the beam or within individual sources. In this case, the RM scatter is more naturally explained by a varying foreground Faraday screen rather than by Faraday complexity intrinsic to the polarized sources. A more detailed separation of these effects would require additional modeling of the polarized spectra, for example through QU-fitting \citep[e.g.,][]{farnsworth_2011,osullivan_2012,miyashita_2019}, which is beyond the scope of this work.

The bridge shows a depolarization ratio comparable to, or slightly higher than, that of the off-target sample. This indicates that the bridge does not introduce strong additional depolarization at these frequencies. Taken together with the small offset suggested by the RM structure function, the results point to weak RM variations on 10 kpc scales in the bridge, with any additional contribution more likely appearing on the larger angular scales probed by the structure function. Deeper RM sampling will be needed to test this trend more robustly.

A possible interpretation is that the magnetic field in the bridge is relatively ordered on these scales, for example if premerger compression aligns the field in the midplane between the clusters and reduces fluctuations in the line-of-sight component. \citet{brzycki2019} provide a qualitative picture of this process using idealized magnetohydrodynamic (MHD) simulations of cluster mergers. Although they focus less on the phase before core passage, they find that the ICM in the midplane between the two clusters becomes compressed and forms a flat, pancake-like structure perpendicular to the line of centers between the cluster cores. The compressed gas amplifies the magnetic field in these regions, and the field appears less turbulent when viewed perpendicular to the merger axis.

The premerger status of the A3391/95 system is supported by recent X-Ray Imaging and Spectroscopy Mission (XRISM) observations. \cite{ota2026} have measured the gas motions in the core region of A3395S using high-resolution X-ray spectroscopy with the Resolve instrument onboard XRISM. By fitting the Fe XXV He alpha complex, they could constrain the line-of-sight bulk and turbulent velocities finding that the one-dimensional turbulent velocity in the central region of A3395S is around 100 km/s. This implies that the ICM is quite calm indicative of a premerger phase. Moreover, they find a line-of-sight bulk velocity of 260 km/s which indicates the presence of large-scale coherent motions in the ICM. They also conclude that a considerable fraction of the merger-driven kinetic energy has not yet been converted into volume-filling turbulence, at least in the X-ray brightest regions.

The RM grid presented here provides a basis for constraining the magnetic-field strength and structure of the merging A3391/95 galaxy-cluster system. However, RM depends on both the line-of-sight magnetic field and the thermal electron-density distribution, so robust quantitative estimates require detailed modeling and comparison with simulations \citep[e.g.,][]{lopez_2025}. Such modeling is beyond the scope of this work, which instead focuses on the observed quantities and on the methodology used to construct the dense RM grid.

\subsection{Upper limit on the radio bridge}
\label{sec:upper_limit}

A non-detection of diffuse radio emission in the X-ray bridge between these two clusters has implications for particle-acceleration mechanisms in large-scale structure. A well-studied intercluster bridge is the A399-A401 system, for which radio, optical and X-ray observations support the scenario where A399–A401 is in the initial phase of merger and the two clusters have not yet started to interact \citep{govoni_2019}. X-ray observations show hot gas not only in the central regions of both clusters, but also in the connecting intercluster region, where the X-ray surface brightness is enhanced. At low radio frequencies, LOFAR images at 144 MHz revealed a bridge of diffuse emission across the intercluster region, implying relativistic electrons and magnetic fields distributed on megaparsec scales \citep{govoni_2019}. Follow-up work has reinforced that the bridge emission is very steep-spectrum. It is not detected at 346 MHz with the Westerbork Synthesis Radio Telescope (WSRT), which implies a steep spectral index \citep{nunhokee_2023, nishiwaki2026} and a recent LOFAR analysis combining 60 and 144 MHz measurements provides a direct spectral constraint on the bridge emission \citep{pignataro_2024}. These results suggest that radio bridges may be most readily detectable at low frequencies, and that a non-detection at GHz frequencies does not by itself exclude faint, steep-spectrum intercluster emission.

In the MeerKAT image, we also find no clear diffuse radio halo emission associated with the central regions of A3391 or A3395. This differs from A399-A401, where both clusters host radio halos in addition to the intercluster bridge \citep{murgia_2010,govoni_2019,nunhokee_2023}. The absence of bright halos in A3391/95 suggests that there is no obvious cluster-scale reservoir of diffuse synchrotron emission connected to the bridge region. This may lower the expectation of detecting bright bridge emission at GHz frequencies, although it does not rule out fainter emission from weak shocks, turbulence, or fossil relativistic plasma in the intercluster medium. Further high-sensitivity observations of A3391/95 at lower frequencies would be needed to test this possibility.

In the MeerKAT Stokes $I$ image, no diffuse synchrotron emission is detected that is spatially coincident with the X-ray bridge between Abell 3391 and Abell 3395. To quantify the corresponding upper limit, we followed the box-statistics approach of \citet{bruggen_2021}. This method is equivalent to measuring the surface brightness after smoothing to the angular scale of the bridge region, while using off-source boxes to estimate the residual background. It therefore improves the sensitivity to very extended emission compared to the per-beam image noise, approximately by averaging over the independent beams contained in the bridge aperture. To amplify diffuse emission on large scales expected by the bridge, we used the compact source subtracted $1\arcmin$ Stokes I image to determine the upper limit. We placed a square box of side length $1120\arcsec$ on the bridge region and compared its mean surface brightness to that measured in independent control boxes of the same size outside the bridge. Any compact radio emission left after source subtraction was excluded from the calculation using a $5\sigma_I$ mask.

The bridge box has a mean surface brightness of $3.48\times10^{-6}$, while the control boxes have a mean surface brightness of $-1.24\times10^{-6}$ Jy beam$^{-1}$ and a box-to-box scatter of $4.70\times10^{-6}$. This corresponds to a residual bridge flux density of only $1.07$ mJy, with an uncertainty of $1.44$ mJy, implying a significance of only $0.74\sigma$ and therefore no detection of diffuse bridge emission. Using a $3\sigma$ upper limit on the diffuse bridge flux density of $4.33$ mJy at the MeerKAT central frequency of 1279 MHz, we derive the corresponding upper limit on the radio emissivity in the bridge region.

For a direct comparison with the ASKAP result of \citet{bruggen_2021}, we scale this limit to 1 GHz assuming a spectral index of $\alpha=-1.3$ \citep{govoni_2019}. This yields a corresponding upper limit of $5.97$ mJy at 1 GHz. Using the same bridge dimensions as in \citet{bruggen_2021}, namely a cylindrical volume with length $1.3$ Mpc and radius $0.6$ Mpc, the corresponding upper limit on the radio emissivity is

\begin{equation}
\begin{aligned}
\langle J \rangle_{1\,\mathrm{GHz}}
&< \frac{4\pi d_L^2 S_{1\,\mathrm{GHz}}}{V} \\
&< \frac{5.97 \times 10^{-29}\,\mathrm{W\,Hz^{-1}\,m^{-2}} \times 4\pi \times 5.4 \times 10^{49}\,\mathrm{m}^2}{4.3 \times 10^{67}\,\mathrm{m}^3} \\
&< 9.21 \times 10^{-46}\,\mathrm{W\,Hz^{-1}\,m^{-3}}.
\end{aligned}
\end{equation}

This limit is more than an order of magnitude lower than the ASKAP-based upper limit reported by \citet{bruggen_2021}, owing to the higher sensitivity of the MeerKAT data.

\section{Conclusions}
\label{sec:conclusions}

We have presented wide-field MeerKAT L-band polarimetry of the Abell 3391/95 premerger system. Using a three-pointing mosaic combined with direction-dependent calibration, visibility plane mosaicing, and Faraday synthesis, we produced high-sensitivity full-Stokes products over a $3\degree \times 2\degree$ field and constructed a dense RM grid to probe the magneto-ionic environment of the system. We identified a total of 434 polarized sources within the field. The polarized source density varies strongly across the mosaic, ranging from about 30 sources per square degree in the outer regions to about 110 sources per square degree in the central part of the field, with a field-averaged density of 73 sources per square degree. Our main results are as follows:

\begin{enumerate}

    \item We detect a statistically significant enhancement of the RM scatter in the clusters relative to the off-target control region. The measured scatters are $\sigma_{\rm RM,on}=29.31 \pm 1.26$ rad m$^{-2}$ and $\sigma_{\rm RM,off}=18.86 \pm 0.69$ rad m$^{-2}$, corresponding to an excess scatter of $22.42 \pm 1.73$ rad m$^{-2}$. The RM-scatter enhancement is driven by the cluster regions. Within $r_{500}$ we measure $\sigma_{\rm RM}=35.90 \pm 4.49$ rad m$^{-2}$ toward A3391 and $\sigma_{\rm RM}=39.31 \pm 2.03$ rad m$^{-2}$ toward the A3395 system, both substantially above the off-target level. This behavior is consistent with Faraday rotation fluctuations produced by turbulent magnetic fields and spatial variations in the thermal electron density in the intracluster medium. These results provide a basis for future quantitative constraints on the magnetic-field strength and structure of the A3391/95 system through detailed modeling and comparison with simulations.

    \item In the intercluster bridge region, the RM scatter is comparatively low, $\sigma_{\rm RM}=13.98 \pm 1.64$ rad m$^{-2}$, below the off-target scatter. Taken at face value, this implies a non-detection of an additional contribution to the RM scatter above the off-target level. However, the bridge occupies a much smaller area near the mosaic center, so the scatter mainly reflects smaller angular scales than the off-target sample. When comparing bridge and off-target regions on matched angular separations using an RM structure function, the bridge shows systematically higher $D_{\rm RM}$ values, although the difference is not statistically significant (one-sided permutation-test p-value of 0.11). One interpretation is that the magnetic field in the bridge is relatively ordered on beam and source scales, for example due to premerger compression that flattens and aligns the field in the mid-plane between the clusters, as predicted by idealized MHD simulations of cluster pairs in the precontact phase. This picture is also consistent with recent XRISM measurements indicating that the ICM in A3395S is dynamically calm, suggesting that merger-driven, volume-filling turbulence has not yet fully developed.

    \item We measured depolarization ratios between the 893.8-1275 MHz and 1282-1663 MHz subbands for polarized sources across the field. The cluster A3391 shows a lower median depolarization ratio ($0.59 \pm 0.02$) than the off-target sample ($0.74 \pm 0.02$), consistent with enhanced Faraday complexity in the cluster environment. The bridge depolarization ratio ($0.78 \pm 0.02$) is comparable to the off-target value, indicating no strong additional depolarization attributable to the bridge at these frequencies, consistent with weak RM variations on small scales and any possible bridge-related contribution being more apparent only on larger angular separations.

    \item No compelling diffuse synchrotron emission is detected in association with the intercluster bridge in our MeerKAT L-band data. After compact-source subtraction and low-resolution imaging, we do not recover additional large-scale diffuse emission across the field.
\end{enumerate}

\section*{Data availability}

The science-ready MeerKAT data products presented in this paper, including the high-resolution Stokes~$I$ mosaic, the $9\arcsec$ $IQUV$ images, the peak polarized-intensity map, and the polarized-source RM catalog, are available at Zenodo at \url{https://doi.org/10.5281/zenodo.21640023}.

\begin{acknowledgements}

VG acknowledges support by the German Federal Ministry of Education and Research (BMBF) under grant D-MeerKAT III. 
We thank Jeremy Sanders for helping us with the eROSITA data.
MB acknowledges funding by the Deutsche Forschungsgemeinschaft (DFG, German Research Foundation) under Germany's Excellence Strategy -- EXC 2121 ``Quantum Universe'' --  390833306 and the DFG research unit "Relativistic Jets".
SPO acknowledges support from the Comunidad de Madrid Atracción de Talento program via grant 2022-T1/TIC-23797, and grant PID2023-146372OB-I00 funded by MICIU/AEI/10.13039/501100011033 and by ERDF, EU.
We acknowledge data storage and computational facilities by the University of Bielefeld which are hosted by the Forschungszentrum Jülich and that were funded by German Federal Ministry of Education and Research (BMBF) projects D-LOFAR IV (05A17PBA) and D-MeerKAT-II (05A20PBA), as well as technical and operational support by BMBF projects D-LOFAR 2.0 (05A20PB1) and D-LOFAR-ERIC (05A23PB1).
The MeerKAT telescope is operated by the South African Radio Astronomy Observatory, which is a facility of the National Research Foundation, an agency of the Department of Science, Technology and Innovation.
We thank the anonymous referee for their constructive comments and suggestions, which have helped to improve the manuscript.

\end{acknowledgements}

\bibliographystyle{aa}
\bibliography{ref}

\begin{appendix}

\section{RM-scatter and depolarization measurements}
\label{app:rm_scatter_depol}

\subsection{RM-scatter excess}

\begin{enumerate}
    \item For each source, we estimated the measurement uncertainty as
    
    \begin{equation}
        \delta \text{RM}_i = \frac{\text{FWHM}}{2 \text{SNR}_i},
    \end{equation}
    
    where \text{FWHM} is the full width at half maximum of the central-facet Faraday depth PSF, which with our weighting scheme is $44.9 \text{ rad m}^{-2}$. The signal-to-noise ratio, $\mathrm{SNR}_i$, is the peak polarized intensity of each source divided by the local noise estimate.

    \item We estimated the uncertainty of the local GRM correction using the standard error of the mean,
    
    \begin{equation}
        \delta \text{GRM}_i = \frac{\sigma_{\text{ann},i}}{\sqrt{N_i}},
    \end{equation}
    
    where $\sigma_{\text{ann},i}$ is the scatter of the annulus RM distribution, measured with the same MAD-based estimator as in the main analysis, and $N_i$ is the number of sources in the annulus.

    \item We then propagated these contributions into the residual RM uncertainty as the quadratic sum,
    
    \begin{equation}
        \delta \text{RRM}_i = \sqrt{(\delta \text{RM}_i)^2 + (\delta \text{GRM}_i)^2}.
    \end{equation}

    \item We propagated the per-source uncertainties to the on-target and off-target scatters, and to the excess RM scatter, using Monte-Carlo realizations. In each realization $k$, we perturbed each RRM value by adding a random deviate drawn from a Gaussian distribution $\mathcal{N}$ with zero mean and standard deviation $\delta \mathrm{RRM}_i$
    
    \begin{equation}
        \text{RRM}^{(k)}_i = \text{RRM}_i + \mathcal{N}\left(0, \delta \text{RRM}_i\right),
    \end{equation}
    
    recomputed $\sigma_{\text{RM,on-target}}^{(k)}$, $\sigma_{\text{RM,off-target}}^{(k)}$, and
    
    \begin{equation}
        \sigma_{\text{RM}}^{\text{excess},(k)} = \sqrt{\left(\sigma_{\text{RM,on-target}}^{(k)}\right)^2 - \left(\sigma_{\text{RM,off-target}}^{(k)}\right)^2},
    \end{equation}
    
    and repeated this procedure for many realizations.
\end{enumerate}

\subsection{Depolarization ratio}
\label{app:depolarization_ratio}

The depolarization analysis was performed using the following procedure:

\begin{enumerate}
\item We divided the data into two subbands spanning 893.8--1275 MHz and 1282--1663 MHz.

\item For each subband, we estimated the per-channel noise in Stokes $Q$ and $U$ from the MAD of all pixels in each image and used these noise estimates to form inverse-variance spectral weights.

\item For each source, and for each subband separately, we extracted the Stokes $I$, $Q$, and $U$ spectra within the source mask.

\item Using the source RM from the full-band analysis, we derotated the $Q$ and $U$ spectra and then formed weighted subband averages.

\item We integrated the total intensity and the Rician-bias-corrected polarized intensity over the source mask to obtain $I_{\rm int}$ and $P_{\rm int}$ in each subband. Sources where the subband Rician bias exceeded the uncorrected polarized intensity were excluded from the analysis.

\item We computed the fractional polarization in each subband,
\begin{equation}
    p = \frac{P_{\rm int}}{I_{\rm int}},
\end{equation}
and defined the depolarization ratio for each source as
\begin{equation}
    \mathrm{DP} = \frac{p_{\rm lo}}{p_{\rm hi}}.
\end{equation}

\item The per-source depolarization uncertainty, $\delta \mathrm{DP}$, was estimated by first-order error propagation,
\begin{equation}
    \delta \mathrm{DP} =
    \mathrm{DP}
    \sqrt{
    \left(\frac{\delta p_{\rm lo}}{p_{\rm lo}}\right)^2
    +
    \left(\frac{\delta p_{\rm hi}}{p_{\rm hi}}\right)^2
    },
\end{equation}
where $\delta p_{\rm lo}$ and $\delta p_{\rm hi}$ are the uncertainties on the low- and high-frequency fractional polarizations, respectively. Assuming that the signal-to-noise in Stokes $I$ is sufficiently high that the Stokes-$I$ errors are negligible, we approximated
\begin{equation}
    \delta p = \frac{\sigma_{QU}}{I},
\end{equation}
where $\sigma_{QU}$ is the propagated noise estimate of the integrated polarized signal in the corresponding subband.

\item Finally, to estimate the median depolarization ratio and its uncertainty for each region, we used Monte Carlo realizations by perturbing each measurement with Gaussian noise of standard deviation $\delta \mathrm{DP}_i$. In each realization $k$, we drew
\begin{equation}
    \mathrm{DP}^{(k)}_i =
    \mathrm{DP}_i +
    \mathcal{N}(0,\delta \mathrm{DP}_i),
\end{equation}
recomputed the regional median depolarization, and summarized the resulting distribution by its median and MAD-based uncertainty.
\end{enumerate}

\section{Polarization of resolved radio sources}
\label{sec:appendix}

In this section we briefly discuss the polarization properties of the radio sources highlighted in Fig.~\ref{fig:StokesI}. We focus on the relation between the total-intensity and polarized-intensity morphologies, identifying regions where the polarized emission closely follows the Stokes $I$ structure and those where it appears more localized, filamentary, or diffuse.

\paragraph{EMU ES J0626–5341 (top left):}

This source shows bright, extended polarized emission distributed along a curved, filamentary ridge. The polarized signal is fairly continuous over much of the Stokes I structure, with several localized peaks embedded in a broader polarized envelope. The polarization follows the total-intensity morphology closely, indicating ordered magnetic-field structure over a large fraction of the source. The source has a mean fractional polarization of 0.61\% and a mean RRM of 35 rad m$^{-2}$.

\paragraph{EMU ES J0632–5404 (top center):}

This source is also strongly polarized and elongated, with polarized emission detected over most of its extent. The brightest polarized regions are concentrated in a few knots along the source, while lower-level polarized emission fills the connecting structure. Overall, the polarized morphology appears coherent and closely associated with the Stokes I emission. The source has a mean fractional polarization of 0.89\% and a mean RRM of $-19$ rad m$^{-2}$.

{\setlength{\textfloatsep}{6pt plus 1pt minus 2pt}%
\setlength{\floatsep}{6pt plus 1pt minus 2pt}%
\setlength{\intextsep}{6pt plus 1pt minus 2pt}%
\setlength{\abovecaptionskip}{4pt}%
\setlength{\belowcaptionskip}{0pt}%

\begin{figure*}[!t]
    \centering

    \includegraphics[width=0.318\textwidth]{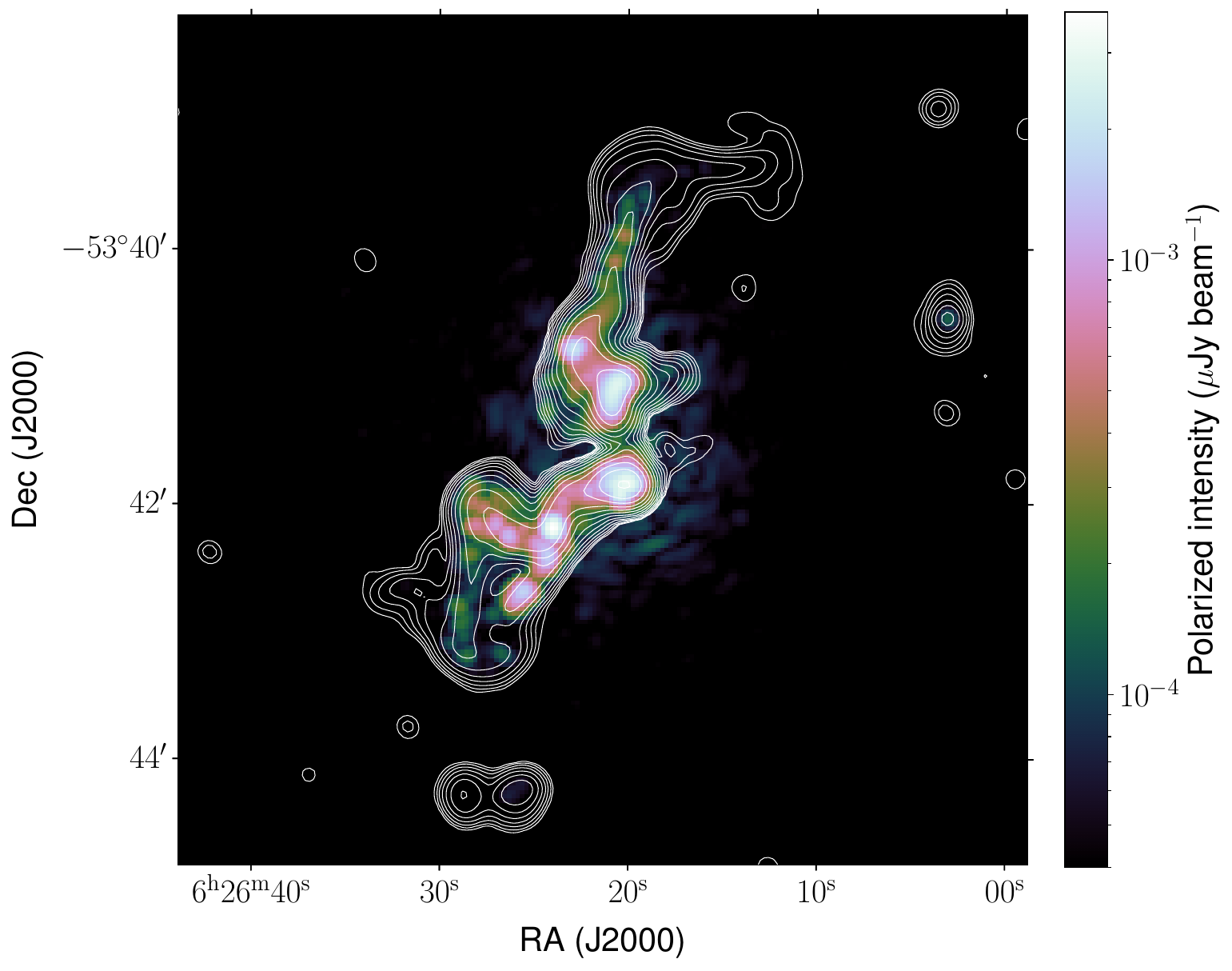}\hfill
    \includegraphics[width=0.318\textwidth]{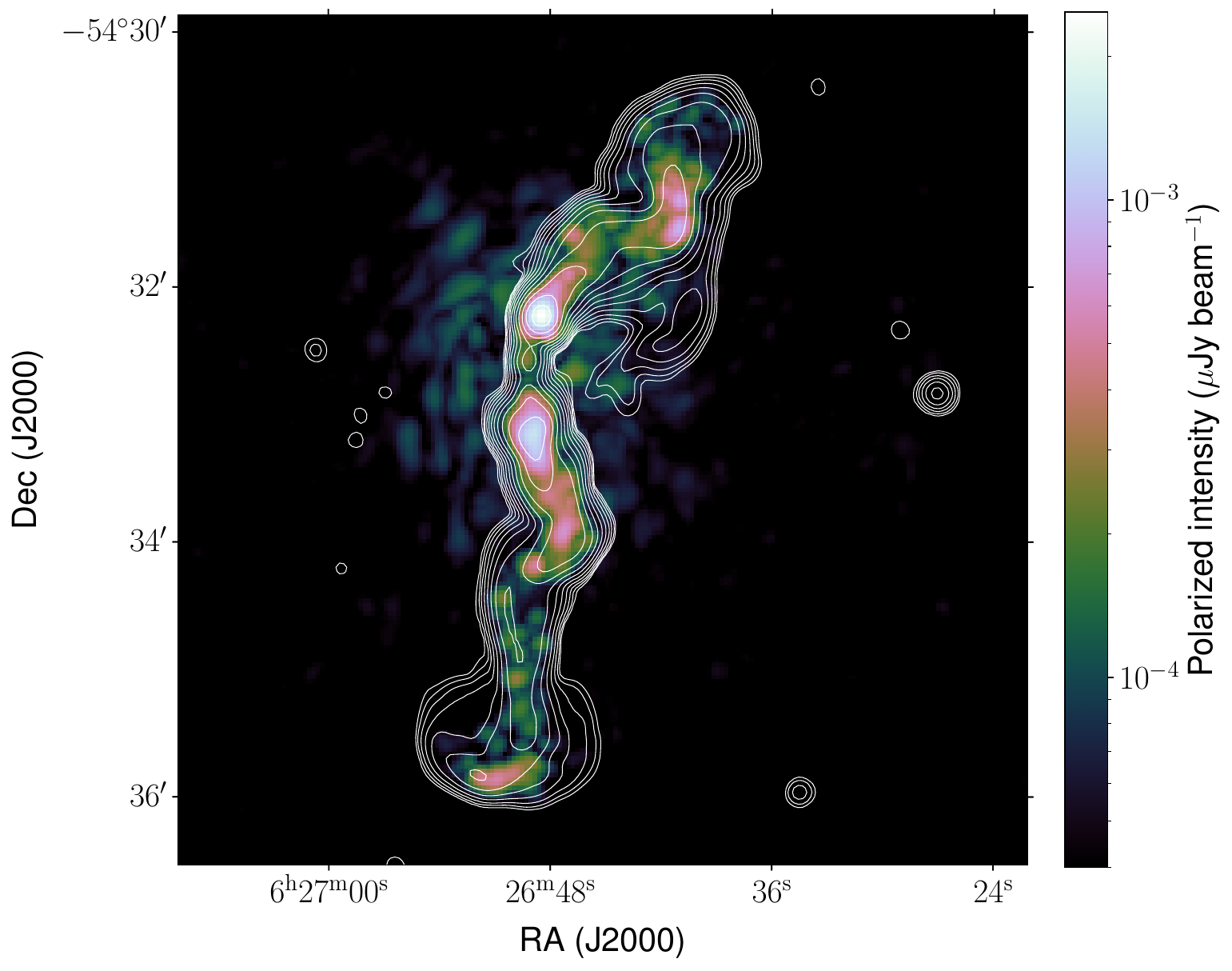}\hfill
    \includegraphics[width=0.318\textwidth]{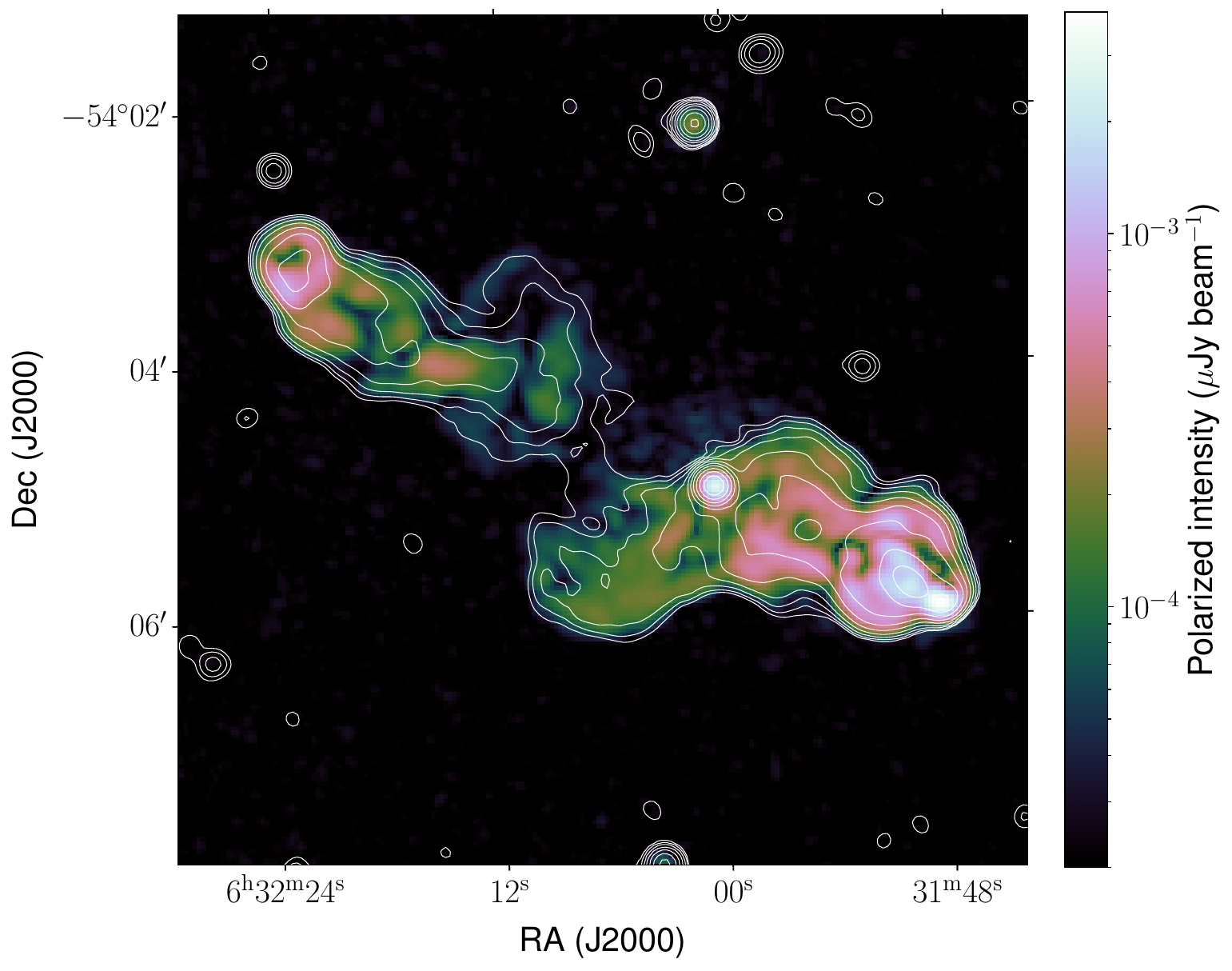}

    \vspace{-0.2em}

    \makebox[\textwidth][c]{%
        \includegraphics[width=0.318\textwidth]{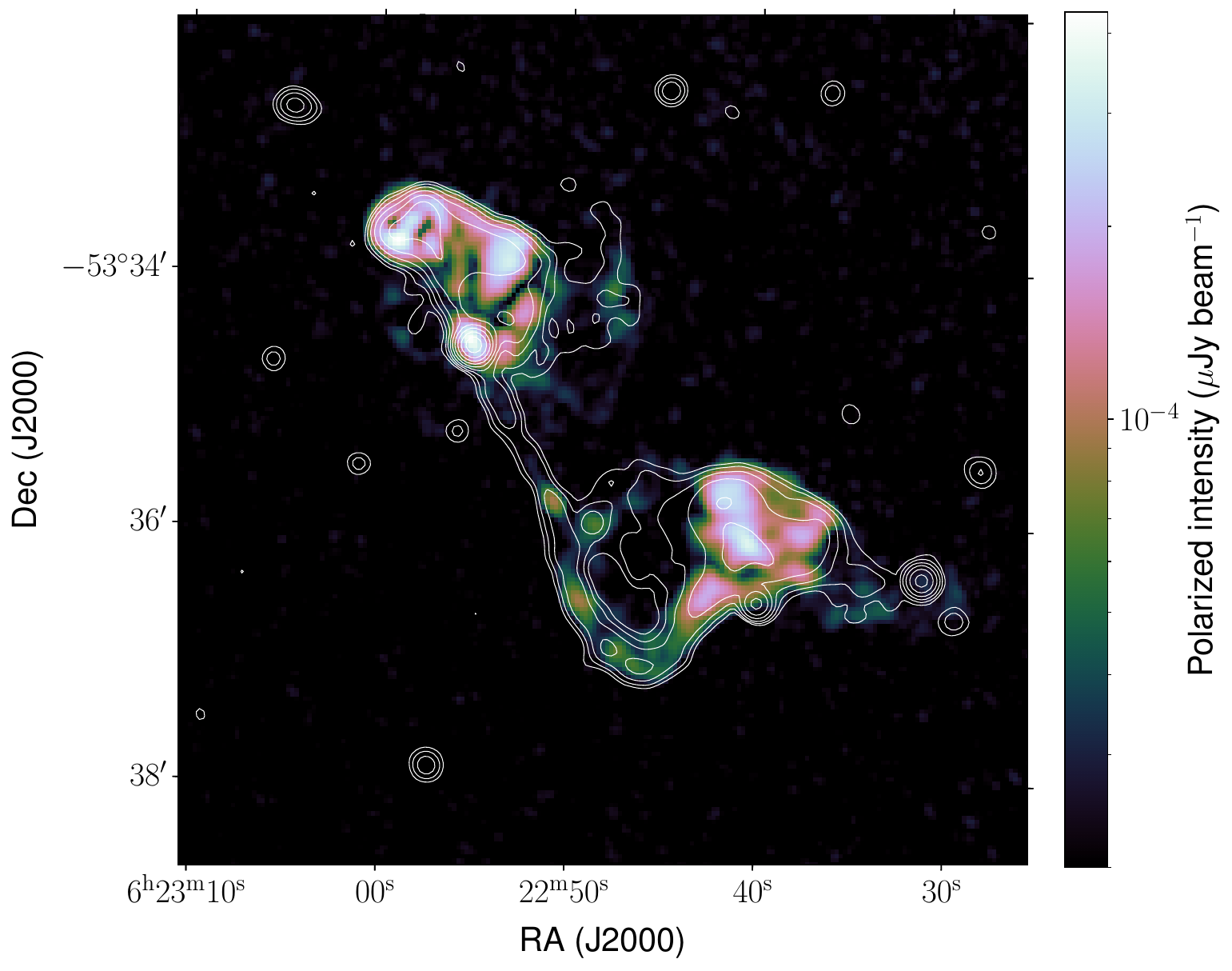}\hspace{0.012\textwidth}%
        \includegraphics[width=0.318\textwidth]{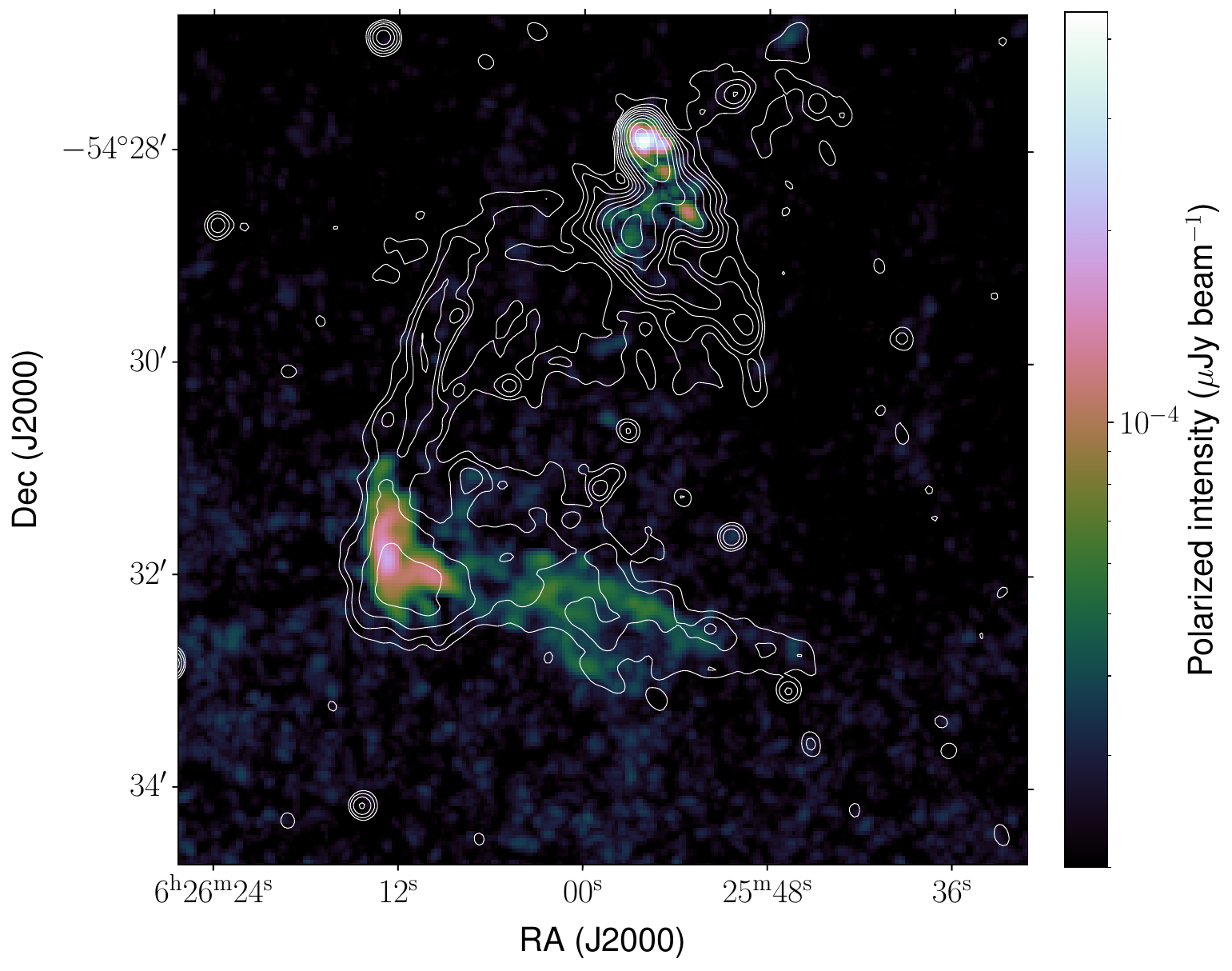}%
    }

    \caption{Rician debiased peak polarized-intensity cutouts of resolved radio sources in the field, with Stokes $I$ contours overlaid in white. Contours start at the local $5\sigma_I$ level and increase by a factor of 2.}
    \label{fig:all_cutouts}
\end{figure*}
}

\paragraph{EMU ES J0626–5432 (top right):}

This source consists of two separated components, both of which show substantial polarized emission. The eastern component is narrower and more filamentary, whereas the western component is broader and hosts the brightest polarized peak. The polarization is clearly detected across both components, but it is not spatially uniform and is instead concentrated in localized substructures. The source has a mean fractional polarization of 14\% and a mean RRM of 22 rad m$^{-2}$.

\paragraph{EMU ES J0622–5334 (bottom left):}

The polarized emission is split into two main lobes connected by a fainter collimated jet. Compared with the first two sources, the polarized morphology is more fragmented and patchy, with the brightest polarized signal concentrated in the outer parts of the two lobes. The jet also shows detectable polarized emission, though at lower surface brightness. The source has a mean fractional polarization of 12\% and a mean RRM of 0.41 rad m$^{-2}$.

\paragraph{2MASX J06261051-5432261 (bottom right):}

This is the most complex source in the field. The polarized emission is extended, irregular, and highly non-uniform, with bright polarized peaks superposed on broader diffuse structure. The morphology suggests that the polarized emission traces not only the radio galaxy associated with 2MASX J06261051-5432261 itself, but also a larger interconnected diffuse radio structure. \cite{bruggen_2021} interprets the emission as consisting of the radio galaxy 2MASX J06261051-5432261 together with additional filamentary and diffuse components, some of which may represent reaccelerated aged plasma rather than emission from a single ordinary radio galaxy alone. The two sources have mean fractional polarizations of 1.3\% and 22\%, and mean RRMs of $-3.4$ rad m$^{-2}$ and 28.6 rad m$^{-2}$ for the northern and southern sources, respectively.

\end{appendix}

\end{document}